\documentclass[MSNbibl,nameyear,dvips]{arxstspdf}
\usepackage{flushend}
\usepackage{stfloats}
\usepackage{graphicx}
\volume{27}
\issue{3}
\pubyear{2012}
\firstpage{350}
\lastpage{372}
\doi{10.1214/11-STS377}

\makeatletter
\newtheorem{theorem}{Theorem}
\newtheorem{proposition}[theorem]{Proposition}

\newcommand{\eqref}[1]{(\ref{#1})}
\renewcommand{\citep}[1]{\citeauthor{#1}, \citeyear{#1}}
\makeatother

\begin{document}
\begin{frontmatter}

\title{Regularization of Case-Specific Parameters for Robustness and Efficiency}
\runtitle{Regularization of Case-Specific Parameters}

\begin{aug}
\author[a]{\fnms{Yoonkyung} \snm{Lee}\corref{}\ead[label=e1]{yklee@stat.osu.edu}},
\author[b]{\fnms{Steven N.} \snm{MacEachern}\ead[label=e2]{snm@stat.osu.edu}}
\and
\author[c]{\fnms{Yoonsuh} \snm{Jung}\ead[label=e3]{yjung1@mdanderson.org}}
\runauthor{Y. Lee, S. N. MacEachern and Y. Jung}

\affiliation{Ohio State University, Ohio State University and University of Texas}

\address[a]{Yoonkyung Lee is Associate Professor, Department of
Statistics, Ohio State University, Columbus, Ohio 43210, USA
\printead{e1}.}
\address[b]{Steven N. MacEachern is Professor, Department of
Statistics, Ohio State University, Columbus, Ohio 43210, USA \printead{e2}.}
\address[c]{Yoonsuh Jung is Postdoctoral Fellow,
University of Texas, MD Anderson Cancer Center,
Houston, Texas 77030, USA \printead{e3}.}

\end{aug}

%
\begin{abstract}
Regularization methods allow one to handle a variety of inferential
problems where there are more covariates than cases. This allows
one to consider a potentially enormous number of covariates for a
problem. We exploit the power of these techniques,
supersaturating models by augmenting the ``natural'' covariates in the problem
with an additional indicator for each case in the data set.
We attach a~penalty term for these case-specific indicators
which is designed to
produce a desired effect. For regression methods with squared error
loss, an~$\ell_1$ penalty produces a regression which is robust to
outliers and high leverage cases; for quantile regression methods,
an $\ell_2$ penalty decreases the variance of the fit enough to
overcome an increase in bias. The paradigm thus allows us to
robustify procedures which lack robustness and to increase the efficiency
of procedures which are robust.

We provide a general framework for the inclusion of case-specific
parameters in regularization problems, describing the impact on the
effective loss for a variety of regression and classification problems.
We outline a computational strategy by which existing software can
be modified to solve the augmented regularization problem, providing
conditions under which such modification will converge to the optimum
solution. We illustrate the benefits of including case-specific parameters
in the context of mean regression and quantile regression through
analysis of NHANES and linguistic data sets.
\end{abstract}

%

\end{frontmatter}

\section{Introduction}\label{sec1}
A core part of regression analysis involves the examination and
handling of individual cases (Weis\-berg, \citeyear{weisberg:book}).
Traditionally, cases have been
removed or downweighted as outliers or because they exert
an overly large influence
on the fitted regression surface. The mechanism by which they
are downweighted or removed is through inclusion of case-specific
indicator variables. For a least-squares fit, inclusion of a
case-specific indicator in the model is equivalent to removing
the case from the data set; for a normal-theory,
Bayesian regression analysis,
inclusion of a case-specific indicator with an appropriate prior
distribution is equivalent to inflating the variance of the
case and hence downweighting it. The tradition in robust regression
is to handle the case-specific decisions automatically, most often
by downweighting outliers according to an iterative procedure
(\citep{huber}).

This idea of introducing
case-specific indicators also applies naturally to criterion based
regression procedures.
Model selection criteria such as AIC or BIC take aim at choosing a model
by attaching a~penalty for each additional parameter in the model.
These criteria can be applied
directly to a larger space of models---namely those in which the
covariates are augmented by a set of case indicators, one for
each case in the data set. When considering inclusion of a case
indicator for a large outlier, the criterion will judge the trade-off
between the empirical risk (here, negative log-likelihood) and model complexity
(here, number of parameters) as favoring the more complex model. It will
include the case indicator in the model, and, with a least-squares
fit, effectively remove the case from the data set. A more considered
approach would allow differential penalties for case-specific indicators
and ``real'' covariates.
With adjustment, one can essentially recover the familiar
$t$-tests for outliers (e.g., \citep{weisberg:book}), either
controlling the error
rate at the level of the individual test or controlling the Bonferroni
bound on the familywise error rate.

Case-specific indicators can also be used in conjunction with
regularization methods such as the LASSO (\citep{lasso}).
Again, care must
be taken with details of their inclusion. If these new covariates are
treated in the same fashion as the other covariates in the problem,
one is making an implicit judgment that they should be penalized in
the same fashion. Alternatively, one can allow a second parameter
that governs the severity of the penalty for the indicators. This
penalty can be set with a view of achieving robustness in the analysis,
and it allows one to tap into a large, extant body of knowledge about
robustness (\citep{huber}).

With regression often serving as a motivating\break theme, a host of
regularization methods for model selection and estimation problems
have been developed.
These methods range broadly across the field of statistics. In addition
to traditional normal-theory linear regression,
we find many methods motivated by a~loss which is composed of a negative
log-likeli\-hood and a penalty for model complexity.
Among these regularization methods are
penalized linear regression methods [e.g., ridge regression (Hoerl and Kennard, \citeyear{ridge})
and the LASSO], regression with a~nonparametric mean function,
[e.g., smoothing\break splines\vadjust{\goodbreak} (\citep{wahba90splines}) and generalized
additive
models (\citep{gam})], and extension to regression with nonnormal error
distributions, name\-ly, generalized linear models (\citep{glm}).
In all of these cases, one can add case-specific indicators along
with an appropriate penalty in order to yield an automated, robust analysis.
It should be noted that, in addition to a different severity for the
penalty term, the case-specific indicators sometimes require a different
form for their penalty term.

A second class of procedures open to modification with case-specific
indicators are those motivated by minimization of
an empirical risk function.
The risk function may not be a negative log-likelihood.
Quantile regression (whether linear or nonlinear) falls into this category,
as do modern classification
techniques such as the support vector machine (\citep{vapnik98}) and
the $\psi$-learner (\citep{psi03}). Many of these procedures
are designed with the robustness of the analysis in mind, often
operating on an estimand defined to be the population-level minimizer of
the risk. The procedures are consistent across a wide variety
of data-generating mechanisms and hence are asymptotically robust. They
have little need of further robustification. Instead, scope for bettering
these procedures lies in improving their finite sample properties.
The finite sample performance of many procedures in this
class can be improved by including case-specific indicators in the problem,
along with an appropriate penalty term for them.

This paper investigates the use of case-specific indicators
for improving modeling and prediction procedures in a regularization
framework.
Section~\ref{sec:robustmodel} provides a formal description
of the optimization problem which arises with the introduction
of case-specific indicators. It also describes a computational
algorithm and conditions that ensure the algorithm will obtain the
global solution to the regularized problem.
Section \ref{sec:reg} explains the
methodology for a~selection of regression methods, motivating particular
forms for the penalty terms. Section \ref{sec:class} describes how the
methodology applies to several classification schemes.
Sections \ref{sec:simulate} and \ref{sec:apply} contain simulation
studies and worked examples. We discuss implications
of the work and potential extensions in Section~\ref{sec:disc}.\looseness=1

\section{Robust and Efficient Modeling Procedures}
\label{sec:robustmodel}

Suppose that we have $n$ pairs of observations denoted by
$(x_i,y_i)$, $i=1,\ldots,n$, for statistical modeling and prediction. Here
$x_i = (x_{i1},\ldots,x_{ip})^\top$ with $p$ covariates and the $y_i$'s are
responses. As in the standard setting of regression and
classification, the $y_i$'s are assumed
to be conditionally independent given the $x_i$'s. In this paper, we take
modeling of the data as a procedure of finding a functional
relationship between $x_i$ and $y_i$, $f(x; \beta)$ with
unknown parameters $\beta\in\Bbb{R}^p$ that is consistent with the data.
The discrepancy or lack of fit of $f$ is measured by a loss function
$\mathcal{L}(y, f(x; \beta))$.
Consider a modeling procedure, say, $\mathcal{M}$ of finding $f$
which minimizes ($n$ times) the empirical risk
\[
R_n(f)=\sum_{i=1}^n \mathcal{L}(y_i, f(x_i; \beta))
\]
or its penalized version,
$ R_n(f)+\lambda J(f) = 
\sum_{i=1}^n \mathcal{L}(y_i,\break f(x_i; \beta))+ \lambda J(f)$,
where $\lambda$ is a positive penalty parameter for balancing the data fit
and the model complexity of $f$ measured by $J(f)$.
A variety of common modeling procedures are subsumed under
this formulation, including ordinary linear regression,
generalized linear models, nonparametric regression,
and supervised learning techniques.
For\break brevity of exposition, we identify $f$ with $\beta$ through a parametric
form and view $J(f)$ as a functional depending on $\beta$.
Extension of the formulation presented in this paper
to a nonparametric function $f$ is straightforward via a~basis expansion.

\subsection{Modification of Modeling Procedures}

First, we introduce case-specific parameters,
$\underline{\gamma}=(\gamma_1,\ldots,\gamma_n)^\top$, for the $n$ observations
by augmenting the covariates with $n$ case-specific indicators.
For convenience, we use $\gamma$ to refer to a generic element of
$\underline{\gamma}$, dropping the subscript. 
Motivated by the beneficial effects of regularization,
we propose a general scheme to modify the modeling procedure
$\mathcal{M}$ using the case-specific parameters $\underline{\gamma}$, to enhance
$\mathcal{M}$ for robustness or efficiency.
Define modification of $\mathcal{M}$ to be the
procedure of finding the original model parameters, $\beta$,
together with the case-specific parameters, $\underline{\gamma}$, that minimize
%
\begin{eqnarray}
\label{eq:robustM}
L(\beta,\underline{\gamma})
&=& \sum_{i=1}^n \mathcal{L}(y_i, f(x_i; \beta)+\gamma_i)
\nonumber
\\[-8pt]
\\[-8pt]
\nonumber
&&{}+ \lambda_\beta J(f)
 + \lambda_\gamma J_2(\underline{\gamma}).
\end{eqnarray}
If $\lambda_\beta$ is zero, $\mathcal{M}$ involves
empirical risk minimization, otherwise penalized risk minimization.
The adjustment that the added case-specific parameters bring to
the loss function $\mathcal{L}(y, f(x; \beta))$ is the same regardless of
whether $\lambda_\beta$ is zero or not.\vadjust{\goodbreak}

In general, $J_2(\underline{\gamma})$ measures the size of
$\underline{\gamma}$. When concerned
with robustness, we often take $J_2(\underline{\gamma}) =
\|\underline{\gamma}\|_1=\sum_{i=1}^n|\gamma_i|$.
A rationale for this choice is that with added flexibility,
the case-specific parameters can curb the undesirable influence of individual
cases on the fitted model.
To see this effect, consider minimizing $L(\hat{\beta}, \underline
{\gamma})$ for
fixed $\hat{\beta}$, which decouples to a~minimization of
$\mathcal{L}(y_i, f(x_i; \hat{\beta})+\gamma_i)
+ \lambda_\gamma|\gamma_i|$ for each $\gamma_i$. In most cases,
an explicit form of the minimizer $\hat{\underline{\gamma}}$ of
$L(\hat{\beta},\underline{\gamma})$ can be obtained.
Generally $\hat{\gamma}_i$'s
are large for observations with large ``residuals'' from the current
fit, and
the influence of those observations
can be reduced in the next round of fitting $\beta$ with the
$\hat{\underline{\gamma}}$-adjusted data.
Such a case-specific adjustment would be necessary
only for a small number of potential outliers, and
the $\ell_1$ norm which yields sparsity works to that effect.
The adjustment in the process of sequential updating of $\beta$
is equivalent to changing the loss from
$\mathcal{L}(y, f(x;\beta))$ to $\mathcal{L}(y, f(x;\beta)+\hat{\gamma})$,
which we call the \textit{$\gamma$-adjusted loss} of $\mathcal{L}$.
The $\gamma$-adjusted loss is a re-expression of $\mathcal{L}$
in terms of the adjusted residual, used as a conceptual aid to illustrate
the effect of adjustment through the case-specific parameter~$\gamma$ on
$\mathcal{L}$.
Concrete examples of the adjustments will be given in the following sections.
Alternatively, one may view
$\mathcal{L}_{\lambda_\gamma}(y, f(x;\beta))
:=\min_{\gamma\in\Bbb{R}} \{ \mathcal{L}(y, f(x; \beta)+\gamma) + \lambda
_\gamma
|\gamma|\} = \mathcal{L}(y, f(x; \beta)+\hat{\gamma})
+ \lambda_\gamma|\hat{\gamma}|$
as a whole to be the ``effective loss'' in terms of $\beta$ after profiling
out $\hat{\gamma}$. The effective loss replaces
$\mathcal{L}(y, f(x; \beta))$ for the modified $\mathcal{M}$ procedure.
When concerned with efficiency, we often take
$J_2(\underline{\gamma}) = \|\underline{\gamma}\|_2^2=\sum_{i=1}^n\gamma_i^2$.
This choice has the effect of increasing the impact of selected,
nonoutlying cases on
the analysis.

In subsequent sections,
we will take a few standard statistical methods for regression and
classification and illustrate how this general scheme applies.
This framework allows us to see established procedures in a new
light and also generates new procedures.
For each method, particular attention will be paid to the form of
adjustment to
the loss function by the penalized case-specific parameters.

\subsection{General Algorithm for Finding Solutions}

Although the computational details for obtaining the solution to
\eqref{eq:robustM} are specific to each modeling procedure $\mathcal{M}$,
it is feasible to describe a common computational strategy which is
effective for a wide range of procedures that optimize a convex function.
For fixed $\lambda_\beta$ and $\lambda_\gamma$, the solution
pair of $\hat{\beta}$ and $\hat{\underline{\gamma}}$ to the modified
$\mathcal{M}$
can be found with little extra computational cost.
A~generic algorithm below alternates estimation of $\beta$ and
$\underline{\gamma}$.\vadjust{\goodbreak}
Given $\hat{\underline{\gamma}}$, minimization of $L(\beta,
\hat{\underline{\gamma}})$ is done via the original modeling procedure
$\mathcal{M}$.
In most cases we consider, minimization of $L(\hat{\beta}, \underline
{\gamma})$
given $\hat{\beta}$ entails simple adjustment of ``residuals.''
These considerations lead to the following iterative algorithm
for finding $\hat{\beta}$ and $\hat{\underline{\gamma}}$:

\begin{enumerate}
\item Initialize $\hat{\underline{\gamma}}^{(0)}=0$ and
$\hat{\beta}^{(0)}=\operatorname{arg} \min_\beta L(\beta,0)$
(the ordinary $\mathcal{M}$ solution).
\item Iteratively alternate the following two steps, $m=0,1,\ldots$:

\begin{itemize}
\item$\hat{\underline{\gamma}}^{(m+1)}
= \operatorname{arg} \min_{\underline{\gamma}\in\Bbb{R}^n}
L(\hat{\beta}^{(m)},\underline{\gamma})$ modifies ``re\-siduals.''

\item$\hat{\beta}^{(m+1)}=\operatorname{arg} \min_{\beta\in\Bbb{R}^p} L(\beta,
\hat{\underline{\gamma}}^{(m+1)})$.
This step\break amounts to reapplying the $\mathcal{M}$ procedure to
$\hat{\underline{\gamma}}^{(m+1)}$-adjusted data although the nature of
the data adjustment would largely depend on~$\mathcal{L}$.
\end{itemize}

\item Terminate the iteration when
$\|\hat{\beta}^{(m+1)}-\hat{\beta}^{(m)}\|^2 < \varepsilon$,
where $\varepsilon$ is a prespecified convergence tolerance.
\end{enumerate}

In a nutshell, the algorithm attempts to find the joint minimizer
$(\beta,\underline{\gamma})$ by combining the minimizers $\beta$ and
$\underline{\gamma}$ resulting from the projected subspaces.
Convergence of the iterative updates
can be established under appropriate conditions.
Before we state the conditions and results for convergence,
we briefly describe implicit assumptions on the loss function and the
complexity or penalty terms, $J(f)$ and $J_2(\underline{\gamma})$.
$\mathcal{L}(y, f(x;\beta))$ is assumed to be nonnegative.
For simplicity, we assume that $J(f)$ of $f(x;\beta)$ depends on~$\beta$ only,
and that it is of the form $J(f)=\|\beta\|_k^k$ and
$J_2(\underline{\gamma})=\|\underline{\gamma}\|_k^k$ for $k \geq1$.
The LASSO penalty has $k=1$ while a ridge regression type
penalty sets $k=2$. Many other penalties of this form for $J(f)$
can be adopted as well to achieve better model selection
properties or certain desirable performance
of $\mathcal{M}$. Examples include those for the elastic net
(Zou and Hastie. \citeyear{elasticnet}), the grouped LASSO (\citep{glasso})
and the hierarchical LASSO (\citep{hlasso}).

For certain combinations of the loss $\mathcal{L}$ and the penalty functionals,
$J(f)$ and $J_2(\underline{\gamma})$, more efficient computational
algorithms can be devised, as in Has\-tie et~al. (\citeyear{svmpath}), \citet{lars} and
\citet{rosset:zhu}.
However, in an attempt to provide a~general computational recipe applicable
to a variety of modeling procedures which can be implemented with simple
modification of existing routines,
we do not pursue the optimal implementation tailored to a specific
procedure in this paper.

Convexity of the loss and penalty terms plays a~primary role
in characterizing\vadjust{\goodbreak} the solutions of the iterative algorithm.
For a general reference to properties of
convex functions and convex optimization,
see \citet{convex}. Nonconvex problems require different
optimization strategies.

If $L(\beta,\underline{\gamma})$ in \eqref{eq:robustM} is continuous and
strictly convex in $\beta$ and $\underline{\gamma}$ for fixed $\lambda
_\beta$ and
$\lambda_\gamma$, the minimizer pair $(\beta,\underline{\gamma})$ in
each step is
properly defined. That is, given $\underline{\gamma}$, there exists a
unique minimizer
$\beta(\underline{\gamma}):=\operatorname{arg} \min_\beta
L(\beta,\underline{\gamma})$, and vice versa.
The assumption that $L(\beta,\underline{\gamma})$ is strictly convex holds
if the loss $\mathcal{L}(y, f(x;\beta))$ itself is strictly convex.
Also, it is satisfied when a convex $\mathcal{L}(y, f(x;\beta))$ is combined
with $J(f)$ and $J_2(\underline{\gamma})$ strictly convex in $\beta$
and $\underline{\gamma}$, respectively.

Suppose that $L(\beta,\underline{\gamma})$ is strictly convex in
$\beta$ and~$\underline{\gamma}$ with a unique minimizer
$(\beta^*,\underline{\gamma}^*)$ for fixed $\lambda_\beta$ and~$\lambda
_\gamma$.
Then, the iterative algorithm gives a sequence of
$(\hat{\beta}^{(m)},\hat{\underline{\gamma}}^{(m)})$ with strictly decreasing
$L(\hat{\beta}^{(m)},\hat{\underline{\gamma}}^{(m)})$. Moreover,
$(\hat{\beta}^{(m)},\hat{\underline{\gamma}}^{(m)})$ converges to
$(\beta^*,\underline{\gamma}^*)$. This result of convergence of the iterative
algorithm is well known in convex optimization, and
it is stated here without proof.
Interested readers can find a formal proof in \citet{regcaseparam}.

\section{Regression}
\label{sec:reg}

Consider a linear model of the form
$y_i=x_i^\top\beta+ \varepsilon_i$. Without loss of generality,
we assume that each covariate is standardized.
Let $X$ be an
$n \times p$ design matrix with $x_i^\top$ in the $i$th row and let
$Y=(y_1,\ldots,y_n)^\top$.

\begin{figure*}
\centering
\begin{tabular}{@{}cc@{}}

\includegraphics{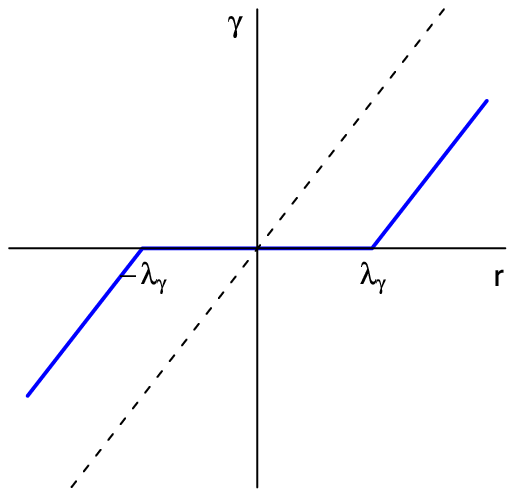}
 & \includegraphics{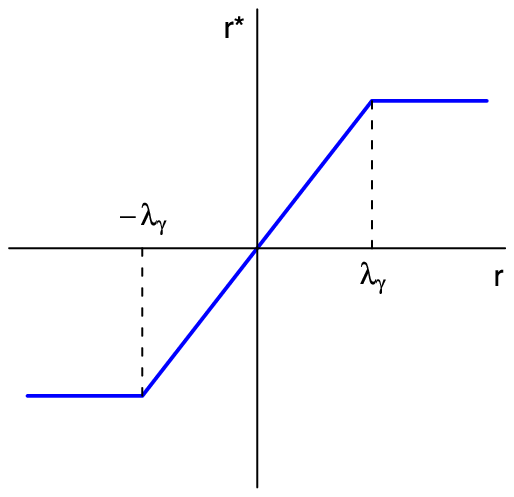}\\
\footnotesize{(a)} & \footnotesize{(b)}\\[6pt]

\includegraphics{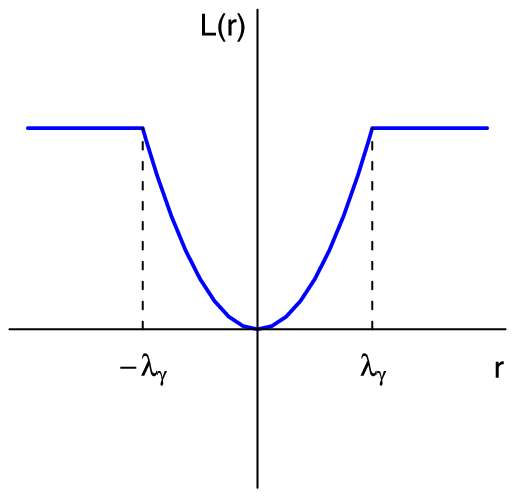}
 & \includegraphics{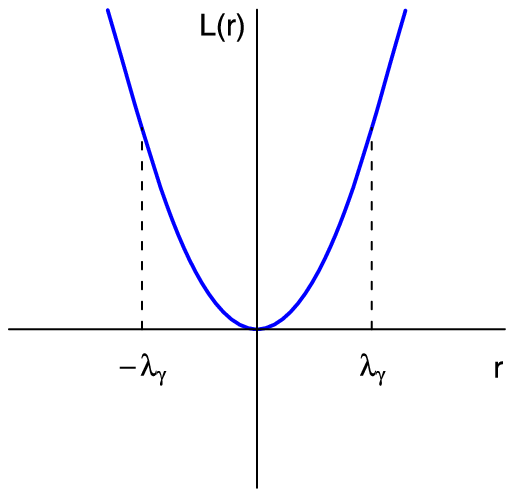}\\
\footnotesize{(c)} & \footnotesize{(d)}
\end{tabular}
\caption{Modification of the squared error loss with a case-specific
parameter.
\textup{(a)} $\gamma$ versus the residual $r$,
\textup{(b)} the adjusted residual $r^*$ versus the ordinary residual $r$,
\textup{(c)} a truncated squared error loss as the $\gamma$-adjusted
loss and \textup{(d)} the effective loss.}
\label{fig:winsorize}
\end{figure*}

\subsection{Least Squares Method}

Taking the least squares method as a baseline modeling procedure
$\mathcal{M}$, we make a link between its modification via
case-specific parameters and a classical robust regression procedure.

The least squares estimator of $\beta=(\beta_1,\ldots,\beta_p)^\top$
is the minimizer $\hat{\beta} \in\Bbb{R}^p$ of
$ L(\beta)=\frac{1}{2} (Y-X\beta)^\top(Y-X\beta)$.
To reduce the sensitivity of the estimator to influential
observations, the $p$ covariates are augmented by $n$ case indicators.
Let $z_i$ be the indicator variable taking 1 for the $i$th observation
and~0 otherwise, and let
$\underline{\gamma}=(\gamma_1,\ldots,\gamma_n)^\top$ be the
coefficients of
the case indicators. The additional design matrix $Z$
for $z_i$ is the identity matrix, and $Z\underline{\gamma}$ becomes
$\underline{\gamma}$ itself.
The proposed modification of the least squares method
with $J_2(\underline{\gamma})=\|\underline{\gamma}\|_1=\sum
_{i=1}^n|\gamma_i|$
leads to a~well-known robust regression procedure.
For the robust modification, we find
$\hat{\beta} \in\Bbb{R}^p$ and $\hat{\underline{\gamma}} \in
\Bbb{R}^n$ that minimize
%
\begin{eqnarray}
\label{eq:robustls}\qquad
L(\beta,\underline{\gamma})
&=&\tfrac{1}{2}\{Y-(X\beta+\underline{\gamma})\}^\top
\{Y-(X\beta+\underline{\gamma})\}
\nonumber
\\[-8pt]
\\[-8pt]
\nonumber
&&{}+ \lambda_{\gamma}\|\underline{\gamma}\|_1,
\end{eqnarray}
where $\lambda_{\gamma}$ is a fixed regularization
parameter constraining $\underline{\gamma}$.
Just as the ordinary LASSO with the~$\ell_1$ norm penalty stabilizes regression
coefficients by\break shrinkage and selection,
the additional penalty in \eqref{eq:robustls} has the same effect
on $\underline{\gamma}$, whose components gauge the extent of case influences.

The minimizer
$\hat{\underline{\gamma}}$ of $L(\hat{\beta},\underline{\gamma})$ for
a fixed $\hat{\beta}$ can be found
by soft-thresholding the residual vector $r=Y-X\hat{\beta}$.
That is, $\hat{\gamma}_i = \operatorname{sgn}(r_i)(|r_i|-\lambda_{\gamma})_+$.
For observations with
small residuals, $|r_i|\le\lambda_{\gamma}$,
$\hat{\gamma}_i$ is set equal to zero with no effect on
the current fit, and for those with large residuals,
$|r_i|> \lambda_{\gamma}$, $\hat{\gamma}_i$ is
set equal to the residual $r_i=y_i-x_i^\top\hat{\beta}$
offset by $\lambda_{\gamma}$ toward zero.
Combining $\hat{\underline{\gamma}}$ with $\hat{\beta}$,
we define the adjusted residuals to be
$r^*_i=y_i-x_i^\top\hat{\beta}-\hat{\gamma}_i$;
that is, $r^*_i=r_i$ if $|r_i|\le\lambda_{\gamma}$, and
$r^*_i=\operatorname{sgn}(r_i)\lambda_{\gamma}$, otherwise.
Thus, introduction of the case-specific parameters along with the
$\ell_1$ penalty on $\underline{\gamma}$ amounts to
winsorizing the ordinary residuals.
The $\gamma$-adjusted loss is equivalent to
truncated squared error loss which is
$(y-x^\top\beta)^2$ if $|y-x^\top\beta|\le
\lambda_{\gamma}$, and is $\lambda_{\gamma}^2$ otherwise.
Figure \ref{fig:winsorize} shows (a) the relationship between
the ordinary residual~$r$ and the corresponding $\gamma$,
(b)~the residual and the adjusted residual $r^*$, (c) the
$\gamma$-adjusted loss as a~function of~$r$, and (d) the effective
loss.\looseness=-1

The effective loss is
$\mathcal{L}_{\lambda_\gamma}(y,x^\top\beta)=(y-x^\top\beta)^2/2$ if
$|y-x^\top\beta|\le\lambda_{\gamma}$, and
$\lambda_{\gamma}^2/2+\lambda_{\gamma}(|y-x^\top\beta|-\lambda_{\gamma})$
otherwise.
This effective loss matches Huber's loss function for robust regression
(\citep{huber}).
As in robust regression, we choose a sufficiently large~$\lambda_{\gamma}$
so that only a modest fraction of the residuals are adjusted.
Similarly, modification of the LASSO as a~penalized regression procedure
yields the Huberized LASSO described by \citet{huberlasso}.\vspace*{2pt}


\subsection{Location Families}\vspace*{2pt}
More generally, a wide class of problems can be cast in the
form of a minimization of
$L(\beta) =\break \sum_{i=1}^n g(y_i - x_i^\top\beta)$
where $g(\cdot)$ is the negative log-likelihood derived from a location
family. The assumption that we have a location family
implies that the negative log-likelihood is a function only
of $r_i=y_i - x_i^\top\beta$.
Dropping the subscript,
common choices for the negative log-likelihood, $g(r)$, include~$r^2$
(least squares, normal distributions) and $|r|$ (least
absolute deviations, Laplace distributions).

Introducing the case-specific parameters $\gamma_i$, we wish to
minimize
\[
L(\beta, \underline{\gamma}) = \sum_{i=1}^n g(y_i - x_i^\top\beta-
\gamma_i) +
\lambda_\gamma\|\underline{\gamma}\|_1.
\]
For minimization with a fixed $\hat{\beta}$,
the next result applies to a broad class of $g(\cdot)$ (but not to
$g(r) = |r|$).

\begin{proposition}
\label{prop:locfam}
Suppose that $g$ is strictly convex with the minimum at 0,\ and
$\lim_{r\rightarrow\pm\infty}g'(r)=\pm\infty$, respectively.
Then,
%
\begin{eqnarray*}
\hat{\gamma}&=&\operatorname{arg}\min_{\gamma\in\Bbb{R}} ~ g(r - \gamma) + \lambda
_\gamma|\gamma|\\
&=&\cases{
r - g'^{-1}(\lambda_{\gamma}),\vspace*{2pt}\cr
\quad\hspace*{11pt}\mbox{for } r > g'^{-1}(\lambda_{\gamma}), \vspace*{2pt}\cr
0, \quad\mbox{for } g'^{-1}(-\lambda_{\gamma})
\leq r \leq g'^{-1}(\lambda_{\gamma}), \vspace*{2pt}\cr
r - g'^{-1}(-\lambda_{\gamma}), \vspace*{2pt}\cr
\quad\hspace*{11pt}\mbox{for } r < g'^{-1}(-\lambda
_{\gamma}).}
\end{eqnarray*}
\end{proposition}

The proposition follows from straightforward algebra. Set the first
derivative of the decoupled minimization equation equal to $0$ and
solve for
$\gamma$. Inserting these values for $\hat{\gamma}_i$ into the equation for
$L(\beta, \underline{\gamma})$ yields
\[
L(\hat{\beta}, \hat{\underline{\gamma}})
 =  \sum_{i=1}^n g(r_i - \hat{\gamma_i}) +
\lambda_\gamma\|\hat{\underline{\gamma}}\|_1.
\]
The first term in the summation can be decomposed into three parts. Large
$r_i$ contribute $g(r_i - r_i +  g'^{-1}(\lambda_{\gamma})) =
g(g'^{-1}(\lambda_{\gamma}))$. Large, negative $r_i$ contribute
$g(g'^{-1}(-\lambda_{\gamma}))$. Those $r_i$ with intermediate values
have $\hat{\gamma}_i = 0$ and so contribute $g(r_i)$. Thus a~graphical
depiction of the $\gamma$-adjusted loss is much like that in Figure
\ref{fig:winsorize},
panel (c), where the loss is truncated above.
For asymmetric distributions (and hence
asymmetric log-likelihoods), the truncation point may differ for positive
and negative residuals. It should be remembered that when $|r_i|$ is
large, the corresponding $\hat{\gamma}_i$ is large, implying a large
contribution of $\|\underline{\gamma}\|_1$
to the overall minimization problem. The residuals will
tend to be large for vectors $\beta$ that are at odds with the data. Thus,
in a sense, some of the loss which seems to disappear due to the effective
truncation of $g$ is shifted into the penalty term for $\gamma$.
Hence the effective loss
$\mathcal{L}_{\lambda_\gamma}(y,f(x; \beta))=g(y-f(x; \beta)-\hat{\gamma})
+ \lambda_\gamma|\hat{\gamma}|$ is the same as the original loss,
$g(y-f(x; \beta))$ when the residual is in
$[g'^{-1}(-\lambda_{\gamma}),\break g'^{-1}(\lambda_{\gamma})]$ and is
linear beyond the interval. The linearized part of $g$ is joined with
$g$ such that $\mathcal{L}_{\lambda_\gamma}$ is differentiable.

Computationally, the minimization of $L(\beta, \hat{\underline{\gamma}})$
giv\-en~$\hat{\underline{\gamma}}$ entails
application of the same modeling procedure $\mathcal{M}$ with $g$
to winsorized pseudo responses $y^*_i=y_i-\hat{\gamma}_i$,
where $y^*_i=y_i$ for $g'^{-1}(-\lambda_{\gamma})
\leq r_i \leq g'^{-1}(\lambda_{\gamma})$,
$y^*_i = g'^{-1}(\lambda_{\gamma})$ for $r > g'^{-1}(\lambda_{\gamma})$,
and $y^*_i = g'^{-1}(-\lambda_{\gamma})$ for
$r < g'^{-1}(-\lambda_{\gamma})$.
So, the $\hat{\underline{\gamma}}$-adjusted data in Step~2 of the main algorithm
consist of $(x_i,y_i^*)$ pairs in each iteration.
A related idea of subsetting data and model-fitting to the subset
iteratively for robustness can be found in the computer vision
literature, the random sample consensus algorithm (Fischler and Bolles, \citeyear{RSC}) for
instance.

\subsection{Quantile Regression}

\begin{figure*}
\centering
\begin{tabular}{@{}ccc@{}}

\includegraphics{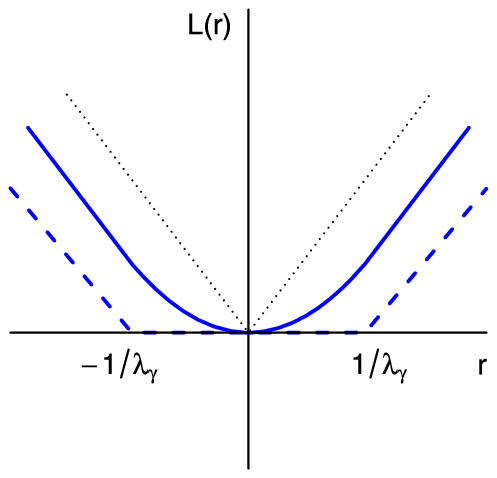}
 & \includegraphics{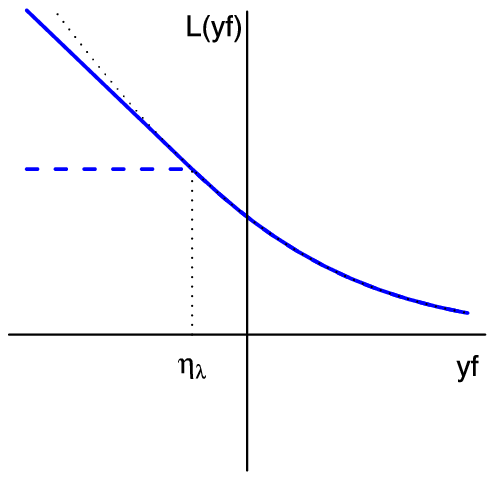} & \includegraphics{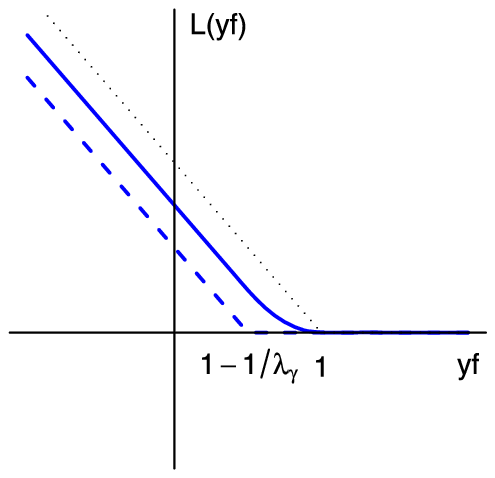}\\
\footnotesize{(a)} & \footnotesize{(b)} & \footnotesize{(c)}
\end{tabular}
\caption{Modification of
\textup{(a)} absolute deviation loss for median regression with $\ell_2$ penalty,
\textup{(b)} negative log-likelihood for logistic regression with $\ell_1$
penalty, and
\textup{(c)} hinge loss for the support vector machine with $\ell_2$ penalty.
The solid lines are for the effective loss, the dashed lines are
for the $\gamma$-adjusted loss, and the dotted lines are
for the original loss in each panel.}
\label{fig:modloss}\vspace*{6pt}
\end{figure*}

Consider median regression with absolute deviation loss
$\mathcal{L}(y,x^\top\beta)=|y-x^\top\beta|$, which is not covered
in the foregoing discussion. It can be verified easily
that the $\ell_1$-adjustment of $\mathcal{L}$ is void due to the
piecewise linearity of the loss, reaffirming the robustness of median
regression.
For an effectual adjustment, the $\ell_2$ norm regularization of
the case-specific parameters is considered.
With the case-specific parameters $\gamma_i$, we
have the following objective function for modified median regression:
%
\begin{equation}
\label{eq:robustmedian}
L(\beta, \underline{\gamma})
=\sum_{i=1}^n |y_i - x_i^\top\beta- \gamma_i|
+ \frac{\lambda_\gamma}{2}\|\underline{\gamma}\|_2^2.
\end{equation}
For a fixed $\hat{\beta}$ and residual
$r=y-x^\top\hat{\beta}$, the $\hat\gamma$ minimizing
$ |r - \gamma|+ \frac{\lambda_\gamma}{2}\gamma^2$
is given by
\[
\operatorname{sgn}(r)\frac{1}{\lambda_\gamma}I \biggl(|r|> \frac{1}{\lambda_\gamma}\biggr)
+r I\biggl(|r|\le\frac{1}{\lambda_\gamma}\biggr).
\]
The $\gamma$-adjusted loss for median regression is
\[
\mathcal{L}(y, x^\top\beta+\hat{\gamma})
=\biggl|y-x^\top\beta-\frac{1}{\lambda_\gamma}\biggr|
I\biggl(|y-x^\top\beta|> \frac{1}{\lambda_\gamma}\biggr),
\]
as shown in Figure \ref{fig:modloss}(a) below.
Interestingly, this $\ell_2$-adjusted absolute deviation loss
is the same as the so-called ``$\varepsilon$-insensitive linear loss''
for support vector regression (\citep{vapnik98}) with
$\varepsilon=1/\lambda_\gamma$.

With this adjustment, the effective loss is Huberized
squared error loss.
The $\ell_2$ adjustment makes
median regression more efficient by rounding the sharp corner of the loss,
and leads to a hybrid procedure which lies between mean and median regression.
Note that, to achieve the desired effect
for median regression, one chooses quite a different value of
$\lambda_\gamma$ than one would when adjusting squared error loss
for a robust mean regression.

The modified median regression procedure can be also combined with
a penalty on $\beta$ for shrinkage and/or selection.
\citet{Bietal} considered support vector regression
with the $\ell_1$ norm penalty $\|\beta\|_1$ for simultaneous robust
regression and variable selection.
These authors relied on the
$\varepsilon$-insensitive linear loss which comes out as the $\gamma$-adjusted
loss of the absolute deviation. In contrast, we rely on the
effective loss which produces a different solution.

In general, quantile regression (Koenker and Bassett, \citeyear{KB1978}; \citep{qr})
can be used to estimate conditional quantiles of $y$ given $x$.
It is a~useful regression technique when
the assumption of normality on the distribution of the errors~$\varepsilon$
is not appropriate, for instance, when the error distribution is skewed or
heavy-tailed. For the $q$th quantile, the check function $\rho_q$
is employed:
%
\begin{equation}
\label{eq:qr}
\rho_q(r) = \cases{
q r, & $\mbox{for }r \ge0,$\vspace*{2pt}\cr
-(1-q)r, & $\mbox{for } r<0.$}
\end{equation}
The standard procedure for the $q$th quantile regression finds $\beta$
that minimizes the sum of asymmetrically weighted absolute errors with
weight $q$ on positive errors and weight $(1-q)$ on negative errors:
\[
L (\beta)= \sum_{i=1}^{n} \rho_q(y_i-x_i^\top\beta).
\]
Consider modification of $\rho_q$ with a case-specific parameter $\gamma
$ and
$\ell_2$ norm regularization. Due to the asymmetry in the loss, except for
$q=1/2$, the size of reduction in the loss by
the case-specific parameter $\gamma$ would depend on its sign.
Given $\hat{\beta}$ and residual $r=y-x^\top\hat{\beta}$,
if $r\ge0$, then the positive $\gamma$ would lower $\rho_q$ by $q\gamma
$, while
if $r<0$, the negative $\gamma$ with the same absolute value would lower
the loss by $(q-1)\gamma$. This asymmetric impact on the loss is
undesirable. Instead, we create a penalty that leads to the same
reduction in loss for positive and negative $\gamma$ of the same
magnitude.
In other words, the desired $\ell_2$ norm penalty needs to
put $q\gamma_+$ and $(1-q)\gamma_-$ on an equal footing. This leads to
the following penalty proportional to $q^2\gamma_+^2$ and
$(1-q)^2\gamma_-^2$:
\[
J_2(\gamma):=\{q/(1-q)\}\gamma_+^2 +\{(1-q)/q\}\gamma_-^2.
\]
When $q=1/2$, $J_2(\gamma)$ becomes the symmetric
$\ell_2$ norm of $\gamma$.

With this asymmetric penalty, given $\hat{\beta}$,
$\hat{\gamma}$ is now defined as
%
\begin{equation}\label{eq:quantile2}
\operatorname{arg} \min_{\gamma\in\Bbb{R}} \mathcal{L}_{\lambda_\gamma}(\hat{\beta},
\gamma)
:=\rho_q(r - \gamma)
+ \frac{\lambda_\gamma}{2} J_2(\gamma),
\end{equation}
and is explicitly given by
\begin{eqnarray*}
&&-\frac{q}{\lambda_\gamma}I \biggl(r<-\frac{q}{\lambda_\gamma}\biggr)
+ r I\biggl(-\frac{q}{\lambda_\gamma}\le r <
\frac{1-q}{\lambda_\gamma}\biggr)
\\
&&\quad{}+\frac{1-q}{\lambda_\gamma}I\biggl(r\ge\frac{1-q}{\lambda_\gamma}\biggr).
\end{eqnarray*}

The effective loss $\rho_q^\gamma$ is then given by
%
\begin{equation}
\label{eq:modcheck}
\rho_q^\gamma(r) = \cases{
(q-1)r-\displaystyle\frac{q(1-q)}{2\lambda_\gamma},\vspace*{2pt}\cr
\quad\mbox{for } r<-\displaystyle\frac{q}{\lambda_\gamma}, \vspace*{2pt}\cr
\displaystyle\frac{\lambda_\gamma}{2} \frac{1-q}{q}r^2,\vspace*{2pt}\cr
\quad\mbox{for } -\!\displaystyle\frac{q}{\lambda_\gamma}\le r < 0, \vspace*{2pt}\cr
\displaystyle\frac{\lambda_\gamma}{2} \frac{q}{1-q}r^2,\vspace*{2pt}\cr
\quad\mbox{for } 0\le r < \displaystyle\frac{1-q}{\lambda_\gamma}, \vspace*{2pt}\cr
qr-\displaystyle\frac{q(1-q)}{2\lambda_\gamma}, \vspace*{2pt}\cr
\quad\mbox{for } r \ge\displaystyle\frac{1-q}{\lambda_\gamma},}
\end{equation}
and its derivative is
%
\begin{equation}
\label{eq:modcheckderiv}
\psi_q^\gamma(r) = \cases{
q-1, & $\mbox{for } \displaystyle r<-\frac{q}{\lambda_\gamma},$ \vspace*{2pt}\cr
\displaystyle\lambda_\gamma\frac{1-q}{q}r,
& $\mbox{for } \displaystyle-\!\frac{q}{\lambda_\gamma}\le r < 0,$ \vspace*{2pt}\cr
\displaystyle\lambda_\gamma\frac{q}{1-q}r,
& $\mbox{for } 0\le r < \displaystyle\frac{1-q}{\lambda_\gamma},$ \vspace*{2pt}\cr
q, & $\mbox{for } r \ge\displaystyle\frac{1-q}{\lambda_\gamma}.$}
\end{equation}
We note that, under the assumption that the density is locally
constant in a neighborhood of the quantile, the quantile remains
the $0$ of the effective $\psi_q^\gamma$ function.

\begin{figure*}

\includegraphics{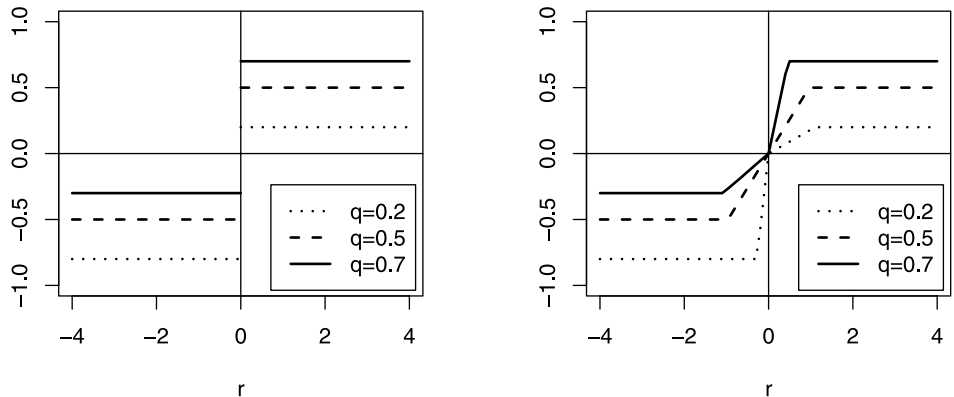}

\caption{The derivative of the check loss
in the left panel, $\psi_q$, and that of the modified check loss
in the right panel, $\psi_q^\gamma$, for $q = 0.2$, $0.5$ and
$0.7$.}\label{fig:psi_fct}\vspace*{-4pt}
\end{figure*}

Figure \ref{fig:psi_fct} compares the derivative of the check loss with
that of the effective loss in \eqref{eq:modcheck}.
Through penalization of a case-specific parameter, $\rho_q$ is
modified to have a continuous derivative at the origin joined by
two lines with a different slope that depends on~$q$.
The effective loss is reminiscent of the
asymmetric squared error loss ($q(r_+)^2 + (1-q)(r_-)^2$)
considered by \citet{NeweyPowell:1987} and \citet{Efron:1991}
for the so-called expectiles.
The proposed modification
of the check loss produces a hybrid of the check loss and asymmetric
squared error loss, however, with different weights than those for
expectiles, to\vadjust{\goodbreak} estimate quantiles.
The effective loss is formally similar to the rounded-corner
check loss of \citet{NychkaEtAl:1995} who used a vanishingly
small adjustment to speed computation. \citet{KoenkerPortnoy:1997}
thoroughly discussed efficient computation for quantile regression.

Redefining $J_2(\underline{\gamma})$ as the sum of the asymmetric~pen\-alty
for the case-specific parameter $\gamma_i$, $i=1,\ldots, n$, modified
quantile regression is formulated as a~procedure that finds
$\beta$ and $\underline{\gamma}$ by minimizing
%
\begin{equation}
\label{eq:qrmodobj}\quad
L(\beta, \underline{\gamma})
=\sum_{i=1}^n \rho_q(y_i - x_i^\top\beta-\gamma_i)
+ \frac{\lambda_\gamma}{2} J_2(\underline{\gamma}).
\end{equation}

In extensive simulation studies (Jung, MacEachern and Lee, \citeyear{l2QR}), such
adjustment of the standard quantile regression procedure generally
led to more accurate estimates.
See Section \ref{subsec:qrsim} for a summary of the studies.
This is confirmed in the\break NHANES data analysis in Section \ref{subsec:nhanes}.

For large enough samples, with a fixed $\lambda_{\gamma}$,
the bias of the enhanced estimator will typically
outweigh its benefits. The natural approach is to adjust the penalty attached
to the case-specific covariates as the sample size increases.
This can be accomplished by increasing the parameter $\lambda_\gamma$
as the sample size $n$ grows.

Let $\lambda_\gamma:=cn^\alpha$ for some constant $c$ and $\alpha>0$.
The following theorem shows that if $\alpha$ is sufficiently large,
the modified quantile regression estimator $\hat{\beta}_n^\gamma$,
which minimizes $\sum_{i=1}^n \rho_q^\gamma(y_i - x_i^\top\beta)$
or equivalent\-ly~\eqref{eq:qrmodobj},
is asymptotically equivalent to the standard estimator $\hat{\beta}_n$.
\citet{Knight:1998} proved the asymptotic normality of the regression quantile
estimator $\hat{\beta}_n$ under some mild regularity conditions.
Using the arguments in \citet{Koenker:2005}, we show that
$\hat{\beta}_n^\gamma$ has the same limiting distribution as $\hat{\beta}_n$,
and thus it is $\sqrt{n}$-consistent if $\alpha$ is sufficiently
large.\vadjust{\goodbreak}

Allowing a potentially different error distribution for each observation,
let $Y_1, Y_2, \ldots$ be independent random variables with c.d.f.'s
$F_1, F_2, \ldots$ and suppose that each $F_i$ has
continuous p.d.f. $f_i$.
Assume that the $q$th conditional quantile
function of $Y$ given $x$ is linear in $x$ and given by $x^\top\beta(q)$,
and let $\xi_i(q):=x_i^\top\beta(q)$.
Now consider the following regularity conditions:

\begin{enumerate}[(C-1)]
\item[(C-1)] $f_i(\xi)$, $i=1,2,\ldots,$ are uniformly bound\-ed away
from 0
and $\infty$ at $\xi_i$.
\item[(C-2)] $f_i(\xi)$, $i=1,2,\ldots,$ admit a first-order Taylor expansion
at $\xi_i$, and $f'_i(\xi)$ are uniformly bound\-ed at $\xi_i$.
\item[(C-3)] There exists a positive definite matrix $D_0$ such that
$ \lim_{n\rightarrow\infty}n^{-1}\sum x_ix_i^\top=D_0$.
\item[(C-4)] There exists a positive definite matrix $D_1$ such that
$ \lim_{n\rightarrow\infty}n^{-1}\sum f_i(\xi_i)x_ix_i^\top=D_1$.
\item[(C-5)] $\max_{i=1,\ldots,n}\| x_i \| /\sqrt{n} \rightarrow0$
in probability.
\end{enumerate}
(C-1) and (C-3) through (C-5) are the conditions considered for
the limiting distribution of the standard regression quantile
estimator $\hat\beta_n$ in \citet{Koenker:2005}
while (C-2) is an additional assumption that we make.

\begin{theorem}
\label{thm:reg_qtl} Under the conditions \textup{(C-1)--(C-5)},
if $\alpha>1/3$, then
\[
\sqrt{n}\bigl(\hat{\beta}^\gamma_{n}-\beta(q)\bigr) \stackrel{d} \rightarrow
N\bigl(0,q(1-q)D_1^{-1}D_0D_1^{-1}\bigr).
\]
\end{theorem}
The proof of the theorem is in the \hyperref[app]{Appendix}.

\section{Classification}
\label{sec:class}

Now suppose that $y_i$'s indicate binary outcomes.
For modeling and prediction of the binary responses, we mainly consider
margin-based procedures such as
logistic regression, support vector machines
(\citep{vapnik98}), and\vadjust{\goodbreak} boosting (\citep{freund:schapire97}). These procedures
can be modified by the addition of case indicators.\vspace*{1pt}

\subsection{Logistic Regression}\vspace*{1pt}
Although it is customary to label a binary outcome as 0 or 1
in logistic regression, we instead adopt the symmetric labels of $\{
-1,1\}$
for $y_i$'s. The symmetry facilitates comparison of different
classification procedures. Logistic regression takes
the negative log-likelihood as a loss for estimation of logit
$f(x)=\log[p(x)/(1-p(x))]$. The loss,
$\mathcal{L}(y,f(x))=\log[1+\exp(-yf(x))]$, can be viewed as a function of
the so-called margin, $yf(x)$. This functional margin of $yf(x)$ is
a pivotal quantity for defining a family of loss functions in classification
similar to the residual in regression.

As in regression with continuous responses, case indicators can be
used to modify the logit function $f(x)$ in logistic regression to minimize
%
\begin{eqnarray}\label{eq:penlogistic}
&& L(\beta_0,\beta, \underline{\gamma})
\nonumber\\
&&\quad= \sum_{i=1}^n \log\bigl(1+\exp\bigl(-y_i\{f(x_i; \beta_0,\beta)+\gamma_i\}
\bigr)\bigr)\\
&&\qquad{}+ \lambda_\gamma\|\underline{\gamma}\|_1,\nonumber
\end{eqnarray}
where $f(x; \beta_0,\beta)=\beta_0 + x^\top\beta$.
When it is clear in context, $f(x)$ will be used as abbreviated notation
for $f(x; \beta_0,\beta)$, a discriminant function, and the
subscript $i$ will be dropped.
For fixed $\hat{\beta}_0$ and $\hat{\beta}$, the minimization
decouples, and $\gamma_i$ is determined
by minimizing
\[
\log\bigl(1+\exp\bigl(-y_i\{f(x_i;\hat{\beta}_0,\hat{\beta})+\gamma_i\}
\bigr)\bigr)+
\lambda_\gamma|\gamma_i|.
\]
First note that the minimizer $\gamma_i$
must have the same sign as $y_i$. Letting $\tau=yf(x)$ and assuming that
$0<\lambda_\gamma<1$, we have
$\arg\min_{\gamma\ge0}
\log(1+\exp(-\tau-\gamma))+ \lambda_\gamma|\gamma|
=\log\{(1-\lambda_\gamma)/\lambda_\gamma\} - \tau$
if $\tau\le\log\{(1-\lambda_\gamma)/\lambda_\gamma\}$, and
0 otherwise.
This yields a truncated negative log-likelihood given by
\[
\mathcal{L}(y,f(x))=\cases{
\log\bigl(1+ \lambda_\gamma/(1-\lambda_\gamma)\bigr),\vspace*{2pt}\cr
\quad \mbox{if }yf(x) \le\log\{(1-\lambda_\gamma)/\lambda_\gamma\},
\vspace*{2pt}\cr
\log\bigl(1+\exp(-yf(x))\bigr), \vspace*{2pt}\cr
\quad \mbox{otherwise,}}
\]
as the $\gamma$-adjusted loss.
This adjustment is reminiscent of
\citeauthor{pregibon82}'s (\citeyear{pregibon82}) proposal
tapering the deviance function so as to downweight extreme observations,
thereby producing a robust logistic regression.
See Figure \ref{fig:modloss}(b) for the $\gamma$-adjusted loss
(the dashed line), where
$\eta_\lambda:=\log\{(1-\lambda_\gamma)/\lambda_\gamma\}$ is
a decreasing function of $\lambda_\gamma$.
$\lambda_\gamma$ determines the level of truncation of the loss.
As $\lambda_\gamma$ tends to $1$, there is no truncation.
Figure \ref{fig:modloss}(b) also shows the effective loss (the solid
line) for
the $\ell_1$ adjustment, which linearizes the negative log-likelihood
below $\eta_\lambda$.\vspace*{1pt}

\subsection{Large Margin Classifiers}\vspace*{1pt}

With the symmetric class labels, the foregoing characterization of
the case-specific parameter $\gamma$ in logistic regression can be
easily generalized to various margin-based classification procedures.
In classification, potential outliers are
those cases with large negative margins.
Let $g(\tau)$ be a loss function of the margin $\tau=yf(x)$. The
following proposition, analogous to Proposition \ref{prop:locfam},
holds for a general family of loss functions.

\begin{proposition}
\label{prop:margin}
Suppose that $g$ is convex and monotonically decreasing in $\tau$,
and $g'$ is continuous.
Then, for $\lambda_{\gamma} < -\lim_{\tau\rightarrow-\infty}g'(\tau)$,
\begin{eqnarray*}
\hat{\gamma}
&=&\arg\min_{\gamma\in\Bbb{R}} ~ g(\tau+ \gamma) + \lambda_\gamma
|\gamma|
\\
&=&\cases{
g'^{-1}(-\lambda_{\gamma})-\tau, & $\mbox{for } \tau
\le g'^{-1}(-\lambda_{\gamma}),$\vspace*{2pt}\cr
0, & $\mbox{for } \tau> g'^{-1}(-\lambda_{\gamma}).$}
\end{eqnarray*}
\end{proposition}

The proof is straightforward. Examples of the mar\-gin-based loss $g$
satisfying the assumption include
the exponential loss $g(\tau)=\exp(-\tau)$ in boosting,
the squared hinge loss $g(\tau)=\{(1-\tau)_+\}^2$ in the support vector
machine (\citep{leemangasarian}), and the negative log-likelihood
$g(\tau)=\log(1+\exp(-\tau))$ in logistic regression.
Although their theoretical targets are different,
all of these loss functions are truncated above for large negative margins
when adjusted by~$\gamma$. Thus, the effective loss
$\mathcal{L}_{\lambda_\gamma}(y, f(x))=g(yf(x)+\hat{\gamma})
+ \lambda_\gamma|\hat{\gamma}|$ is obtained by linearizing
$g$ for $yf(x)<g'^{-1}(-\lambda_{\gamma})$.

The effect of $\hat{\gamma}$-adjustment
depends on the form of~$g$, and hence on the classification
method. For boosting, $\hat{\gamma}\! =\! -\log\lambda_{\gamma}-yf(x)$ if
$yf(x)\!\le\!-\log\lambda_{\gamma}$,~and is 0 otherwise.
This gives
$L(\beta_0, \beta,\hat{\underline{\gamma}})
\!=\!\sum_{i=1}^n \exp(-y_i\cdot f(x_i;\beta_0, \beta)-\hat{\gamma}_i)
\!=\!\sum_{i=1}^n \exp({-}\hat{\gamma}_i) \exp({-}y_if(x_i; \beta_0,\beta)\!)$.
So, finding $\beta_0$ and $\beta$ given $\hat{\underline{\gamma}}$
amounts to weighted boosting, where
the positive case-specific parameters $\hat{\gamma}_i$ downweight the
corresponding cases by\break $\exp(-\hat{\gamma}_i)$.
For the squared hinge loss in the support vector machine, $\hat{\gamma}
= 1-yf(x)-\lambda_{\gamma}/2$ if $yf(x) \le1-\lambda_{\gamma}/2$,
and is
$0$ otherwise.
A positive case-specific parameter $\hat{\gamma}_i$
has the effect of relaxing the margin requirement, that is,
lowering the joint of the hinge for that specific case.
This allows the associated slack variable to be smaller in the primal
formulation. Accordingly, the adjustment affects the coefficient of the linear
term in the dual formulation of the quadratic programming problem.

As a related approach to robust classification,
\citet{Wu:Liu} proposed truncation of margin-based loss functions
and studied theoretical properties that ensure classification
consistency. Similarity exists between our proposed adjustment of a loss
function with $\gamma$ and truncation of the loss at some point.
However, it is the linearization of a margin-based loss function
on the negative side that produces its effective loss, and
minimization of the effective loss is quite different from
minimization of the truncated (i.e., adjusted) loss.
Linearization is more conducive to computation
than is truncation. Application of the result
in \citet{bjm-ccrb-05} shows that the linearized loss
functions satisfy sufficient conditions for classification consistency,
namely Fisher consistency, which is the main property investigated by
\citet{Wu:Liu} for truncated loss functions.

\citet{Xuetal:2009} showed that regularization in the standard
support vector machine is equivalent to a robust formulation
under disturbances of $x$ without penalty. In contrast, under our
approach, robustness of classification methods is considered through
the margin, which is analogous to the residual in regression.
This formulation can cover outliers due to perturbation in $x$ as
well as mislabeling of~$y$.

\subsection{Support Vector Machines}

As a special case of a large margin classifier,
the linear support vector machine (SVM) looks for the optimal hyperplane
$f(x;\beta_0,\beta)=\beta_0 + x^\top\beta=0$ minimizing
%
\begin{equation}\label{eq:svm2}
\hspace*{19pt} L_{\lambda}(\beta_0,\beta)
= \sum_{i=1}^n [ 1 - y_i f(x_i;\beta_0,\beta) ]_+
+ \frac{\lambda}{2}\|\beta\|_2^2,
\end{equation}
where $[t]_+ = \max(t, 0)$ and $\lambda>0$ is a regularization parameter.
Since the hinge loss for the SVM, $g(\tau)=(1-\tau)_+$, is piecewise linear,
its linearization with $\|\underline{\gamma}\|_1$ is void,
indicating that it has little need of further robustification.
Instead, we consider
modification of the hinge loss with $\|\underline{\gamma}\|_2^2$.
This modification is expected to improve efficiency,
as in quantile regression.

Using the case indicators $z_i$ and their coefficients~$\gamma_i$,
we modify \eqref{eq:svm2}, arriving at the problem of minimizing
%
\begin{eqnarray}\label{eq:modsvm2}
L(\beta_0, \beta, \underline{\gamma})
&=& \sum_{i=1}^n [ 1 - y_i \{f(x_i;\beta_0,\beta)+\gamma_i\} ]_+
\nonumber
\\[-8pt]
\\[-8pt]
\nonumber
&&{}+ \frac{\lambda_\beta}{2}\|\beta\|_2^2
+ \frac{\lambda_\gamma}{2}\|\underline{\gamma}\|_2^2.
\end{eqnarray}
For fixed $\hat{\beta}_0$ and $\hat{\beta}$, the minimizer
$\hat{\underline{\gamma}}$
of $L(\hat{\beta}_0,\hat{\beta},\underline{\gamma})$ is obtained by
solving the decoupled optimization problem of
\[
\min_{\gamma_i \in\Bbb{R}}
[1-y_if(x_i;\hat{\beta}_0,\hat{\beta})-y_i\gamma_i]_+
+ \frac{\lambda_\gamma}{2}\gamma_i^2 \quad\mbox{for each }\gamma_i.
\]
With an argument similar to that for logistic regression,
the minimizer $\hat{\gamma}_i$ should have the same sign as~$y_i$.
Let $\xi\!=\!1-yf$. A simple calculation shows that\looseness=-1
\begin{eqnarray*}
&&\arg\min_{\gamma\ge0}  [\xi-\gamma]_+ + \frac{\lambda_\gamma
}{2}\gamma^2
\\
&&\quad=\cases{
0, & $\mbox{if }\xi\le0,$ \vspace*{2pt}\cr
\xi, & $\mbox{if } 0< \xi< 1/\lambda_\gamma,$\vspace*{2pt}\cr
1/\lambda_\gamma, & $\mbox{if } \xi\ge1/\lambda_\gamma.$}
\end{eqnarray*}\looseness=0
Hence, the increase in margin $y_i\hat{\gamma}_i$ due to inclusion of
$\gamma$ is given by
\begin{eqnarray*}
&&\{1-y_if(x_i)\}I\biggl(0< 1-y_if(x_i) < \frac{1}{\lambda_\gamma}\biggr)\\
&&\quad{}+
\frac{1}{\lambda_\gamma} I\biggl(1-y_if(x_i)\ge\frac{1}{\lambda_\gamma}\biggr).
\end{eqnarray*}
The $\gamma$-adjusted hinge loss is
$\mathcal{L}(y,f(x))=[1-1/\lambda_\gamma-yf(x)]_+$ with the hinge lowered
by $1/\lambda_\gamma$ as shown in Figure \ref{fig:modloss}(c)
(the dashed line).
The effective loss (the solid line in the figure)
is then given by a smooth function with
the joint replaced with a quadratic piece between $1-1/\lambda_\gamma$
and 1
and linear beyond the interval.

\section{Simulation Studies}
\label{sec:simulate}

We present results from various numerical experiments
to illustrate the effect of the proposed modification of modeling procedures
by regularization of case-specific parameters.

\subsection{Regression}
\subsubsection{\texorpdfstring{$\ell_2$-adjusted quantile regression}{l2-adjusted quantile regression}}
\label{subsec:qrsim}

The effectiveness of the $\ell_2$-adjusted quantile regression
depends on the penalty parameter $\lambda_\gamma$ in \eqref{eq:modcheck},
which yields $(-q/\lambda_\gamma, (1-q)/\lambda_\gamma)$
as the interval of quadratic adjustment.

We undertook extensive simulation studies (available in \citep{l2QR})
to establish guidelines for selection of the penalty
parameter~$\lambda_\gamma$ in the linear regression model setting. The studies encompassed
a range of sample sizes, from $10^{2}$ to $10^{4}$, a variety of
quantiles, from $0.1$ to $0.9$,
and distributions exhibiting symmetry, varying degrees of asymmetry,
and a variety of tail behaviors.
The modified quantile regression method was directly implemented by specifying
the effective $\psi$-function~$\psi_q^\gamma$,
the derivative of the effective loss, in the
\texttt{rlm} function in the R package.

An empirical rule was established via
a (robust) regression analysis. The analysis considered
$\lambda_\gamma$ of the form
$c_q n^\alpha/\hat{\sigma}$, where
$c_q$ is a constant depending on $q$ and $\hat{\sigma}$ is a robust
estimate of the scale of the error distribution.
The goal of the analysis was to find $\lambda_\gamma$ which, across a
broad range of conditions, resulted in an MSE near the condition-specific
minimum MSE.
Here MSE is defined as mean squared error of estimated regression
quantiles at a new $X$ integrated over the distribution of the covariates.

\begin{figure*}

\includegraphics{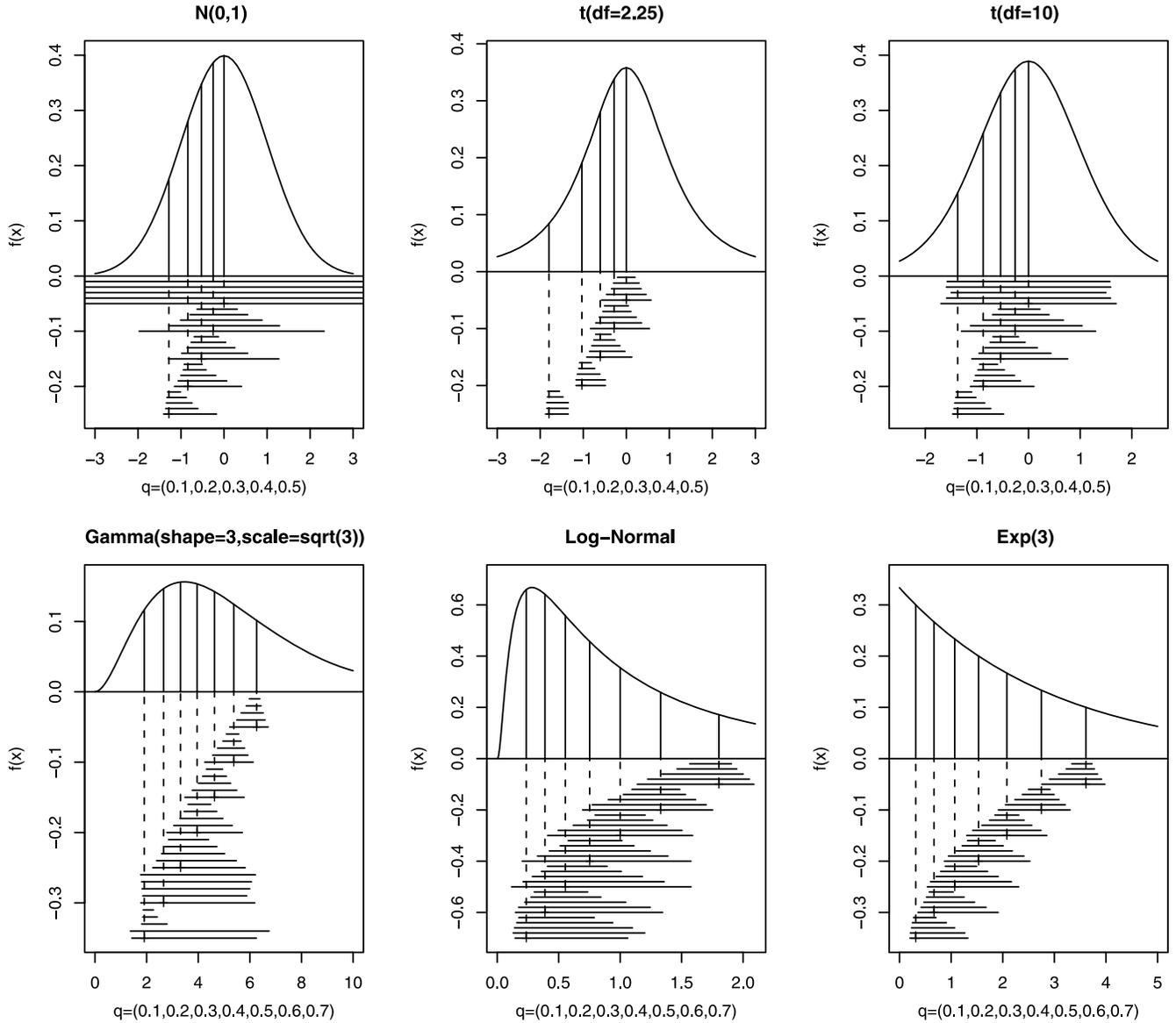}

\caption{``Optimal'' intervals of adjustment for different quantiles ($q$),
sample sizes ($n$), and error distributions.
The intervals range from the quantile minus $q/\lambda_{\gamma}$ to
the quantile plus $(1-q)/\lambda_{\gamma}$, with $\lambda_{\gamma}$
minimizing \textit{MSE}.
The vertical lines in each distribution
indicate the true quantiles. The stacked horizontal
lines for each quantile are corresponding optimal intervals.
Five intervals at each quantile are for $n=10^2$, $10^{2.5}$,
$10^3$, $10^{3.5}$ and $10^4$, respectively, from the bottom.}
\label{fig:window}
\end{figure*}

After initial examination of the MSE with a range of $\alpha$
values, we made a decision to set $\alpha$ to $0.3$ for good finite sample
performance across a wide range of conditions.
With fixed $\alpha$, we varied $c_q$ to obtain the smallest MSE
by grid search for each condition under consideration.
For a quick illustration, Figure \ref{fig:window} shows the intervals
of adjustment with such optimal $c_q$ for various error distributions,
$q$ values, and sample sizes.
Wider optimal intervals indicate that
more quadratic adjustment is preferred to the standard quantile
regression for reduction of MSE.
Clearly, Figure \ref{fig:window} demonstrates the benefit of the proposed
quadratic adjustment of quantile regression in terms of MSE across
a broad range of situations, especially when the sample size is small.

In general, MSE values begin to decrease as the size of adjustment increases
from zero and increase after hitting the minimum, due to an increase
in bias. There is an exception of this typical pattern when estimating
the median with normally distribut\-ed errors. MSE monotonically
decreases in this case as the interval of adjustment widens,
confirming the optimality properties of least
squares regression for normal theory regression.
The comparisons between sample mean and sample median can
be explicitly found under the $t$ error distributions using different
degrees of freedom. The benefit of the median relative to the mean is
greater for thicker tailed distributions.
We observe that this qualitative behavior carries over to the optimal
intervals. Thicker tails lead to shorter optimal intervals, as shown
in Figure \ref{fig:window}.

Modeling the optimal condition-specific $c_q$ as a~function of $q$
through a robust regression analysis led to the rule, with $\alpha
=0.30$, of
$c_q \approx0.5\exp(-2.118-1.097q)$ for $q<0.5$ and
$c_q \approx0.5\exp(-2.118-1.097(1-q))$ for $q \ge0.5$.
The simulation studies show that this choice of penalty parameter
results in an accurate estimator of the quantile surface.

%
\begin{table*}[b]
\tabcolsep=0pt
\caption{Distribution of difference in the number of selected variables
for the fitted model to contaminated data from that
to clean data}\label{tab:simulation1}
\begin{tabular*}{\textwidth}{@{\extracolsep{\fill}}l ccccccc c ccccccc@{}}
\hline
& \multicolumn{7}{c}{\textbf{LARS}} && \multicolumn{7}{c@{}}{\textbf{Robust LARS}}\\
\ccline{2-8,10-16}
\textbf{Scenario} & $\bolds{-3}$ & $\bolds{-2}$ & $\bolds{-1}$ & \textbf{0} & \textbf{1} & \textbf{2} & \textbf{3} &
& $\bolds{-3}$ & $\bolds{-2}$ & $\bolds{-1}$ & \textbf{0} & \textbf{1} & \textbf{2} & \multicolumn{1}{c@{}}{\textbf{3}}\\
\hline
$\varepsilon$ contamination \\
Sparse & 5* & \phantom{0}6 & 21 & 48 & 13 & 5 & \phantom{0}2* &
& 1* & 4 & 12 & 71 & \phantom{0}7 & \phantom{0}5 & 0\\
Intermediate & 5\phantom{*} & 10 & 14 & 46 & 21 & 3 & 1 &
& 1\phantom{*} & 3 & 14 & 64 & 14 & \phantom{0}4 & 0 \\
Dense & 2\phantom{*} & \phantom{0}1 & 16 & 80 & \phantom{0}1 & 0 & 0 &
& 0\phantom{*} & 0 & \phantom{0}8 & 89 & \phantom{0}3 & \phantom{0}0 & 0 \\[3pt]
$X$ contamination \\
Sparse & 7* & \phantom{0}5 & 15 & 34 & 20 & 7 & 12* &
& 5* & 3 & 16 & 36 & 22 & 12 & 6 \\
Intermediate & 1* & \phantom{0}5 & 13 & 55 & 21 & 3 & 2 &
& 1\phantom{*} & 3 & 18 & 50 & 23 & \phantom{0}4 & 1\\
Dense & 0\phantom{*} & \phantom{0}0 & \phantom{0}5 & 93 & \phantom{0}2 & 0 & 0 &
& 0\phantom{*} & 0 & \phantom{0}4 & 94 &\phantom{0}2 & \phantom{0}0 & 0\\
\hline
\end{tabular*}
\tabnotetext[]{}{\textit{Note}:
The entries with * are the cumulative counts of the specified
case and more extreme cases.}
\end{table*}
%

\subsubsection{Robust LASSO}

We investigated the sensitivity of the LASSO
(or LARS) and its robust version (obtained by the
proposed $\ell_1$ modification)
to contamination of the data through simulation.

For the robust LASSO, the iterative algorithm in Section
\ref{sec:robustmodel} was implemented
by using LARS (Efron et~al., \citeyear{lars}) as the baseline modeling procedure
and winsorizing the residuals with $\lambda_{\gamma}$ as a bending
constant. The bending constant was taken to be scale invariant, so
that\vadjust{\goodbreak}
$\lambda_{\gamma} = k \hat{\sigma}$, where $k$ is a constant and
$\hat{\sigma}$ is a~robust scale estimate.
The standard robust statistics literature (\citep{huber})
suggests that good choices of $k$ lie in the range from $1$ to $2$.

For brevity, we report only that portion of the results pertaining to accuracy
of the fitted regression surface and inclusion of variates in the model
when $k=2$.
Similar results were obtained for $k$ near $2$. The results differ
for extreme values of $k$.
Throughout the simulation,
the standard linear model $y= x^\top\beta + \varepsilon$ was assumed.
Following the simulation setting in \citet{lasso},
we generated $x=(x_1,\ldots,x_8)^\top$ from a multivariate normal distribution
with mean zero and standard deviation 1.
The correlation between $x_i$ and $x_j$
was set to $\rho^{|i-j|}$ with $\rho=0.5$.
Three scenarios were considered with a varying degree of sparsity
in terms of the number of nonzero true coefficients:
(i)~sparse: $\beta=(5,0,0,0,0,0,0,0)$,
(ii) intermediate: $\beta=(3,1.5,0,0,2,0,0,0)$ and
(iii) dense: $\beta_j=0.85$ for all $j=1,\ldots,8$.
In all cases, the sample size was $100$.
For the base case, $\varepsilon_i$ was assumed to follow
$N(0,\sigma^2)$ with $\sigma=3$.
For potential outliers in~$\varepsilon$,
the first 5\%
of the $\varepsilon_i$'s were tripled, yielding a data set with more outliers.
We also investigated sensitivity to high leverage cases. For this setting,
we tripled the first 5\% of the values of $x_1$. Thus the replicates
were blocked across the three settings.
The $C_p$ criterion was used to select the model.


\begin{figure}

\includegraphics{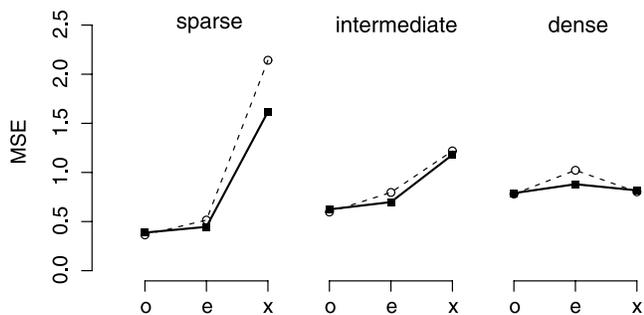}

\caption{Mean squared error (\textit{MSE}) of $\hat\beta$ for LARS and
its robust version under three different scenarios in the simulation study.
In each scenario, \textsf{o}, \textsf{e}, and \textsf{x} indicate
clean data, data with contaminated measurement errors, and
data with mismeasured first covariate.
The dotted lines are for LARS while
the solid lines are for robust LARS.
The points are the average \textit{MSE} for 100 replicates.}
\label{fig:mse.beta}\vspace*{3pt}
\end{figure}

Figure \ref{fig:mse.beta} shows mean squared error (MSE) between the
fitted and true regression surfaces, omitting intercepts.
MSE is integrated across the distribution of a
future $X$, taken to be that for the base case of the simulation. Over the
$m=100$ replicates in the simulation,
$\operatorname{MSE} = m^{-1} \sum_{i=1}^m (\hat{\beta}^i-\beta)^\top
\Sigma(\hat{\beta}^i-\beta)$,
where $\hat{\beta}^i$ is the estimate of $\beta$ for the $i$th
replicate, and $\Sigma$ is the covariance matrix of $X$.
LARS and robust LARS perform comparably in the base case, with the MSE
for robust LARS being greater by $1$ to $6$ percent.
For both LARS and robust LARS, MSE in the base case increases as one moves
from the sparse to the dense scenario. MSE increases noticeably when
$\varepsilon$ is contaminated, by a factor of $1.31$ to $1.41$ for LARS.
For robust LARS, the factor for increase over the base case with LARS is
$1.12$ to $1.22$.
For contamination in $X$, results under LARS and robust LARS are
similar in the
intermediate and dense cases, with increases in MSE over the base
case. For
the sparse case, the coefficient of the contaminated covariate, $x_1$,
is large
relative to the other covariates. Here, robust LARS performs noticeably better
than LARS, with a smaller increase in MSE.

Table \ref{tab:simulation1} presents results on the difference in
number of
selected variables for pairs of models. In each pair, a contaminated model
is contrasted with the corresponding uncontaminated model.
The top half of the table presents results for contamination of
$\varepsilon$.
The distribution of the differences in the number of selected variables
for the pairs of fitted models has a mode at $0$ in each scenario for both
LARS and robust LARS. There is, however,
substantial spread around $0$. The fitted models for the data
with contaminated errors tend to have fewer variables than those
for the original data, especially in the dense scenario. This may well be
attributed to inflated estimates of $\sigma^2$ used in $C_p$ for the
contaminated data, favoring relatively smaller models. The effect is
stronger for LARS than for robust LARS, in keeping with the lessened
impact of outliers on the robust estimate of $\sigma^2$.

The bottom half of Table \ref{tab:simulation1} presents results for
contamination of $X$. Again, the distributions of differences in model size
have modes at $0$ in all scenarios.
The distributions have substantial spread around
$0$. Under the sparse scenario in which the contamination has a substantial
impact on MSE, the distribution under robust LARS is more
concentrated than
under LARS.

The simulation demonstrates that the proposed robustification is
successful in dealing\vadjust{\goodbreak} with both contaminated errors and contaminated
covariates. As expected, in contrast to LARS, robust LARS
is effective in identifying observations with
large measurement errors and lessening their influence. It is also
effective at reducing the impact of high leverage cases, especially when
the high leverage arises from a covariate with a large regression coefficient.
The combined benefits of robustness to outliers and high leverage cases render
robust LARS effective at dealing with influential cases in an automated
fashion.

\subsection{Classification}

A three-part simulation study was carried out to examine the effect of the
proposed modification of loss functions for classification.
The primary focus is on (i)~the efficiency of the modified SVM relative
to the SVM with hinge loss
and its smoothed version with quadratically modified hinge loss,
and (ii) the robustness of logistic regression relative to modified
logistic regression (via the linearized deviance). The secondary
focus is on ensuring that robustness does not significantly degrade
as efficiency is improved, and that efficiency does not suffer
too much as robustness is improved.

All three parts of the simulation begin with $n=100$ cases
generated from a pair of five-dimensional multivariate normal
distributions, with identical covariance matrices and equal proportions
for two clas\-ses ($y=\pm1$). Without loss of generality,
the covariance matrices were taken to be the identity.
For the first part of the simulation, the separation between the
two classes is fixed. The separation is determined by the
difference in means of the multivariate normals, which, in turn,
determine the Bayes error rate for the underlying problem. Throughout,
once a~method was fit to the data (i.e., a discriminant function
was obtained), the error rate was calculated analytically. Each
part of the simulation consisted of $400$ replicates.

Six methods were considered in this study: LDA (linear discriminant
analysis) as a baseline method for the normal setting, the standard
SVM, its variant with squared hinge loss
(called Smooth SVM in \citep{leemangasarian}),
another variant with quadratically modified hinge loss
(referred to as Huberized SVM in \citep{rosset:zhu}),
logistic regression, and the method with linearized binomial deviance
(referred to as linearized LR in this study).
The Huberized SVM and linearized LR were implemented through
the fast Newton--Armijo algorithm proposed\vadjust{\goodbreak} for Smooth SVM
in \citet{leemangasarian}.
To focus on the effect of the loss functions on the classification
error rate,
no penalty was imposed on the parameters of discriminant functions.

\begin{figure}

\includegraphics{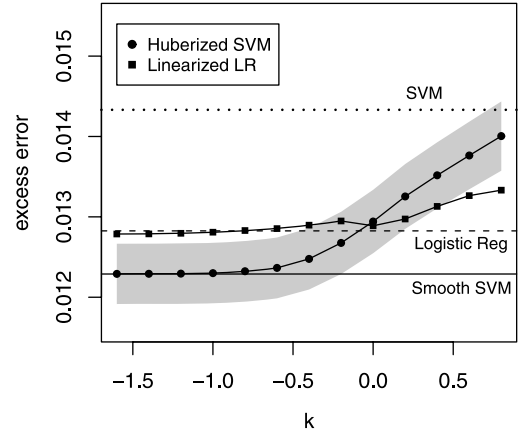}

\caption{Mean excess error of the SVM variant with quadratically
modified hinge loss (Huberized SVM) and the method with
linearized deviance loss (linearized LR)
as the bending constant $k$ varies.
The gray band indicates a one standard error bound
around the mean estimate for Huberized SVM from 400 replicates.
The standard error for comparison of the Huberized SVM to another method
varies, but is considerably smaller, due to the simulation design.
The horizontal lines from top to bottom are for SVM, logistic
regression and Smooth SVM, respectively.}\vspace*{-2pt}
\label{fig:svmvshuberwithk}
\end{figure}

For the first part of the study, the mean vectors were set with a
difference of
2.7 in the first coordinate and 0 elsewhere, yielding a Bayes
error rate of 8.851\%.
Figure \ref{fig:svmvshuberwithk} compares the SVM and its variants
in terms of the average excess error from the Bayes error rate.
The $k$ on the $x$-axis corresponds to
the bending constant, $1-1/\lambda_\gamma$ in the Huberized SVM.
When $k$ is as small as $-1$, we see that
quadratic modification in the Huberized SVM effectively yields the
same result as Smooth SVM. As $k$ tends to 1, the Huberized SVM
becomes the standard SVM. Clearly, there is a range of $k$ values
for which the mean error rate of the Huberized SVM is lower than that
of the standard SVM, demonstrating improved efficiency in
classification. In fact, the improved efficiency of smooth versions of the
hinge loss in the normal setting can be verified theoretically
for large sample cases, where the relative efficiency is defined as
the ratio of mean excess errors. See \citet{lee:wang} for details.

\begin{figure*}
\centering
\begin{tabular}{@{}cc@{}}

\includegraphics{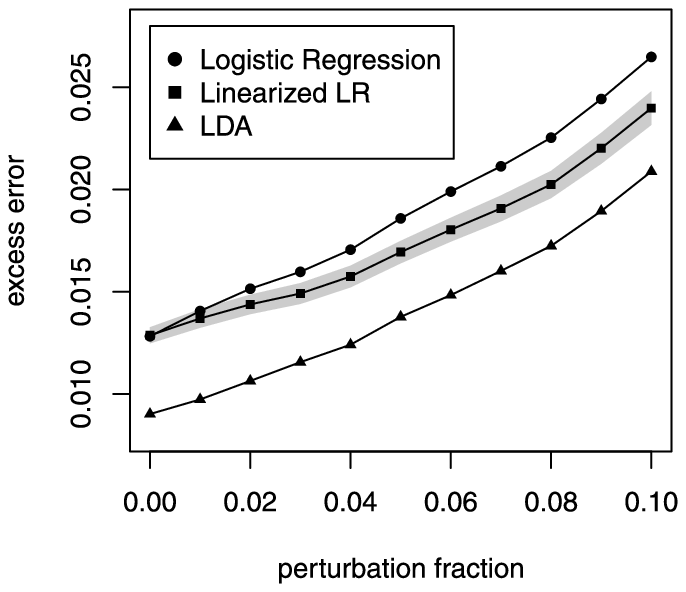}
 & \includegraphics{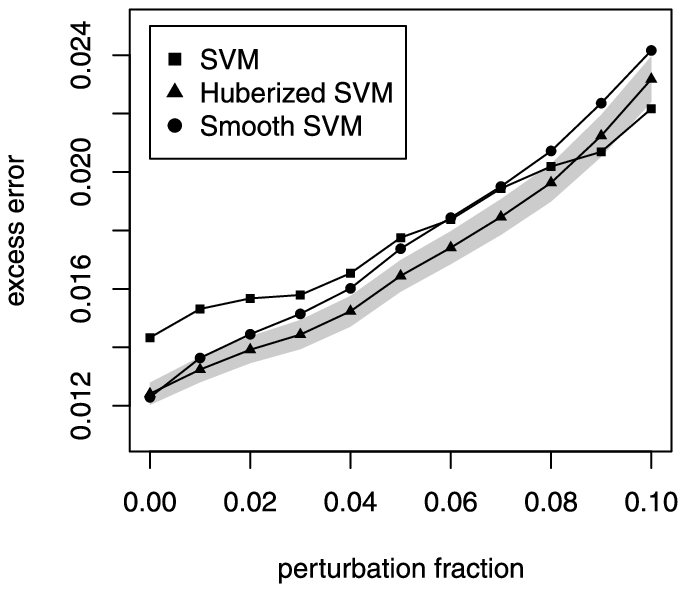}\\
\footnotesize{(a)} & \footnotesize{(b)}
\end{tabular}
\caption{Mean excess errors of \textup{(a)} logistic regression and linearized LR,
and \textup{(b)} SVM and its variants, as the proportion of perturbation varies.
The gray band indicates a one standard error bound
around the mean estimate for \textup{(a)} linearized LR and \textup{(b)} Huberized SVM
from 400 replicates. The standard errors for comparisons are
considerably smaller than indicated by the bands.}
\label{fig:robustclass}
\end{figure*}

Figure \ref{fig:svmvshuberwithk} also displays a comparison
between logistic regression and the linearized LR of
Section \ref{sec:class}, with
bending constant
$k = \log\{(1-\lambda_\gamma)/\lambda_\gamma\}$.
There is no appreciable difference in the excess error between
logistic regression and its linearized version for negative\vadjust{\goodbreak} values of $k$.
Enhancing the robustness of logistic regression (shown in part two of the
study) sacrifices almost none of its efficiency.

The value of the bending constant $k$ leading to the minimum
error rate depends on the underlying problem itself,
and the range of best $k$ values may differ for the
Huberized SVM and linearized LR.
The results in Figure \ref{fig:svmvshuberwithk} suggest that
values of $k$ ranging from $-1$ to $0$ yield excellent
performance for both procedures in this setting.

The second part of the study focuses on robustness. To study this,
we perturbed each sample by flipping the
class labels of a certain proportion of cases selected at random,
and applied the six procedures to the perturbed sample. The
estimated discriminant rules were evaluated in the same
way as in the setting without perturbation.

Figure \ref{fig:robustclass}(a) highlights increased robustness of
linearized LR (with $k=-0.5$) compared to logistic regression when
some fraction of labels are flipped.
As the proportion of mislabeled data increases,
excess error rises for all of the procedures, including
the baseline method of LDA. However, the rate of increase in error
is slower for the modified logistic regression, as the linearized
deviance dampens the influence of mislabeled cases on
the discriminant rule.

Comparison of the SVM and its variants in the same setting
reveals a trade-off between efficiency and robustness.
Figure \ref{fig:robustclass}(b) shows that the squared
hinge loss yields a lower error rate than hinge loss
when the perturbation fraction is less than 6\%. The trend is reversed
when the fraction is higher than 6\%. This trade-off is\vadjust{\goodbreak}
reminiscent of that between the sample mean and median
as location parameter estimators. The Huberized SVM (with $k=-0.5$)
as a hybrid method strikes a balance between the two.
We note that the robustness of the SVM, compared with its variants,
is more visible when two
classes have less overlap (not shown here).

%
\begin{table*}
\tabcolsep=0pt
\caption{Mean error rates of classification methods under various
settings of mean difference and perturbation fraction.
The lowest error rates are in bold when compared among the
methods of the same type (either SVM or LR) for each scenario}\label{tab:error}
\begin{tabular*}{\textwidth}{@{\extracolsep{\fill}}lccccccc@{}}
\hline
&  & \multicolumn{2}{c}{\textbf{Huberized SVM}} &
& \multicolumn{2}{c}{\textbf{Linearized LR}} &  \\
\ccline{3-4,6-7}
\textbf{Scenario} & \textbf{SVM}& $\bolds{k=-0.5}$ & $\bolds{k=-1}$ &\textbf{Smooth SVM} & $\bolds{k=-0.5}$ & $\bolds{k=-1}$ &\textbf{LR}\\
\hline
Easy
& 0.0385 & \textbf{0.0376} & \textbf{0.0376} & \textbf{0.0376}
& \textbf{0.0362} & 0.0363 & 0.0363 \\
Intermediate
&0.1028 & 0.1009 & \textbf{0.1008} & \textbf{0.1008}
& 0.1014 & \textbf{ 0.1013} & \textbf{0.1013} \\
Hard & 0.1753 & 0.1727 & \textbf{0.1726} & \textbf{0.1726}
& 0.1730 & 0.1729 & \textbf{0.1728}\\[3pt]
Easy${} + {}$5\% flip & \textbf{0.0348} & 0.0362 & 0.0371 & 0.0372
& \textbf{0.0383} & 0.0395 & 0.0411 \\
Intermediate${} + {}$5\% flip
& 0.1063 & \textbf{0.1050} & 0.1057 & 0.1059
& \textbf{0.1054} & 0.1061 & 0.1071\\
Hard${} + {}$5\% flip &0.1790 & \textbf{0.1769} & 0.1773 & 0.1774
& \textbf{0.1772} & 0.1773 & 0.1778 \\[3pt]
Easy${} + {}$10\% flip
& \textbf{0.0370} & 0.0415 & 0.0423 & 0.0421
& \textbf{0.0445} & 0.0465 & 0.0481 \\
Intermediate${} + {}$10\% flip
& \textbf{0.1107} & 0.1117 & 0.1127 & 0.1127
& \textbf{0.1125} & 0.1136 & 0.1150 \\
Hard${} + {}$10\% flip
& 0.1846 & \textbf{0.1833} & 0.1839 & 0.1840
& \textbf{0.1836} & 0.1841 & 0.1848 \\
\hline
\end{tabular*}
\end{table*}

The third part of the study provides a comprehensive comparison
of the methods.
Three scenarios with differing degree of difficulty were considered;
``easy,'' ``intermediate'' and ``hard'' settings refer to the multivariate
normal setting with the Bayes error rates of
2.275\%, 8.851\% and 15.866\%, respectively. In addition,
for scenarios with mislabeled cases, 5\% and 10\% of labels were flipped
under each of the three settings. Two values of the bending constant
($k=-0.5$ and $-1$) were used for the Huberized SVM and the linearized LR.
The results of comparison under nine scenarios are
summarized in Table \ref{tab:error}. The tabulated values are the mean
error rates of the discriminant rules under each method.

When there are no mislabeled cases, the smooth variants of the SVM
improve upon the performance of the standard SVM.
As the separation between classes increases, the reduction in
error due to modification of the hinge loss with fixed $k$
diminishes. Linearization of deviance in logistic regression does not
appear to affect the error rate.
In contrast, when there are mislabeled cases, linearization of
the deviance renders logistic regression more robust across all the
scenarios with differing class separations.
Similarly, the standard SVM is less sensitive to mislabeling than
its smooth variants. This makes the SVM more preferable as the
proportion of mislabeled cases increases.
However, in the difficult problem of little class
separation,
the quadratic modification in the Huberized SVM performs better
than the SVM.

\section{Applications}
\label{sec:apply}
\subsection{Analysis of the NHANES Data}
\label{subsec:nhanes}

We numerically compare standard quantile regression with modified
quantile regression for analysis of real data.
The Centers for Disease Control and Prevention conduct the
National Health and Nutrition Examination Survey (NHANES),
a large-scale survey designed to monitor the health and nutrition of
residents of the United States. Many are concerned about the record
levels of obesity in the population, and the survey contains
information on height and weight of individuals, in addition to
a variety of dietary and health-related questions. Obesity is
defined through body mass index (BMI) in $\mathrm{kg}/\mathrm{m}^2$, a measure which adjusts
weight for height. In this analysis,
we describe the relationship between height and BMI
among the $5938$ males over the age of $18$ in the aggregated
NHANES data sets from 1999, 2001, 2003 and 2005. Our analyses
do not adjust for NHANES' complex survey design. In particular,
no adjustment has been made for oversampling of selected groups
or nonresponse.
Since BMI is weight adjusted for height, the null expectation
is that BMI and height are unrelated.

We fit a nonparametric quantile regression model to the data.
The model is a six-knot regression spline using the natural basis expansion.
The knots (held constant across quantiles) were chosen by eye.
The rule for\vadjust{\goodbreak} selection of the penalty parameter $\lambda_\gamma$
described in Section \ref{subsec:qrsim}
was used for the NHANES data analysis.

\begin{figure*}[t]

\includegraphics{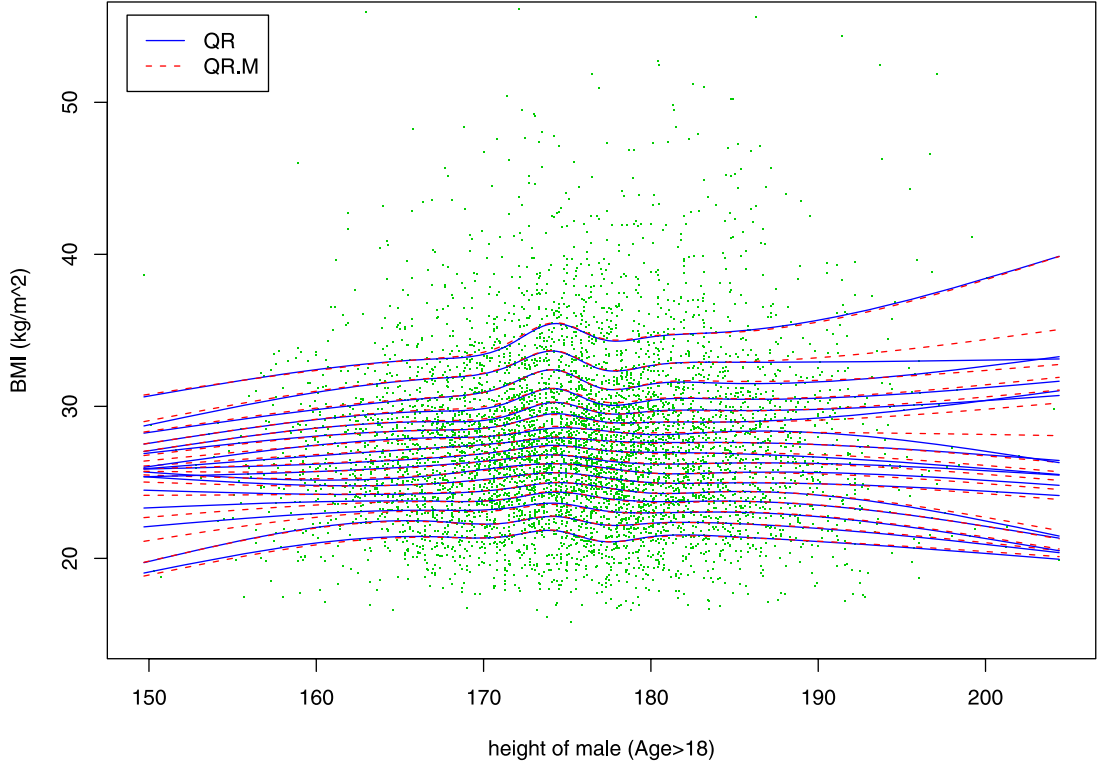}

\caption{Regression spline estimates of conditional BMI quantiles in steps
of $0.05$, from $0.1$ to $0.9$ for the NHANES data. Natural spline
bases and six knots are used in each fitted curve.} \label{fig:NHANES}
\end{figure*}

Figure \ref{fig:NHANES} displays the fits from standard (QR)
and modified (QR.M) quantile regressions for the quantiles between $0.1$
and $0.9$ in steps of $0.05$.
The fitted curves show a slight
upward trend, some curvature overall, and mildly increasing spread
as height increases. There is a noticeable bump upward in the
distribution of BMI for heights near $1.73$ meters.
The differences between the two methods of fitting the quantile
regressions are most apparent in the tails, for example the $0.6$th
and $0.85$th quantiles for large heights.


\begin{figure*}[b]

\includegraphics{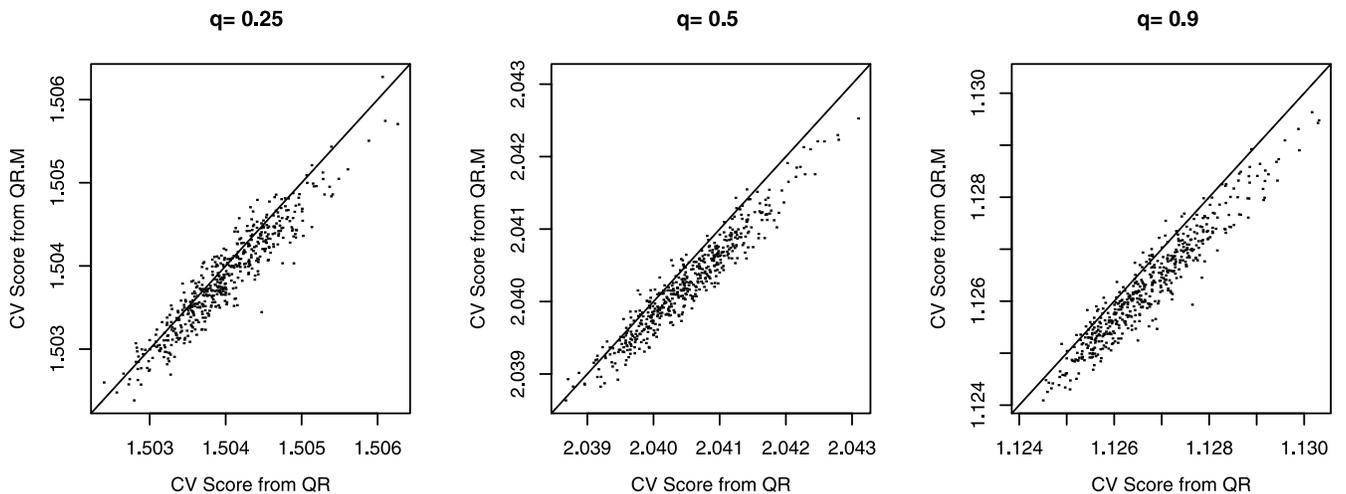}

\caption{Scatterplots of 10-fold CV scores from standard quantile
regression (QR) and modified quantile regression (QR.M) at $0.25$th,
$0.5$th and $0.9$th quantiles.
Regression splines with natural
spline bases and six knots are fitted to the NHANES data. Each of 500
points represents a pair of CV scores as in (\protect\ref{eq:cv}).}
\label{fig:NHANES_MSE_male_df7}
\end{figure*}

The predictive performances of the standard and modified
quantile regressions are compared in
Figure~\ref{fig:NHANES_MSE_male_df7}. To compare the methods,
$10$-fold cross-valida\-tion was repeated $500$ times for different
splits of
the data. Each time, a cross-validation score was computed as
%
\begin{equation}
\label{eq:cv} \mathrm{CV}=
\frac{1}{n}\sum_{i=1}^{n}\rho_q(y_i-\hat{y_i}),
\end{equation}
where $y_i$ is the observed BMI for an individual in the
hold-out sample, $\hat{y}_i$ is the fitted value under QR or QR.M,
and the sum runs over the hold-out sample. The figure contains
plots of the $500$ $\mathrm{CV}$ scores. The great majority of $\mathrm{CV}$ scores are
to the lower right side of the 45 degree line, indicating that
the modified quantile regression outperforms the standard method---even
when the QR empirical risk function is used to evaluate performance.
Mean and 1000 times standard deviation of the $\mathrm{CV}$ scores for the
methods are
summarized in Table~\ref{tab:NHANES}.

The pattern shown in these panels
is consistent across other quantiles (not shown here). The pattern
becomes a bit stronger when the QR.M empirical risk function is
used to evaluate performance.

Quantile regression has the property that
$100\cdot q$\% of the responses fall at or below the fitted $q$th quantile
surface. This does not have to hold for the modified quantile
regression fit.
However, as the cross-validation shows, QR.M does provide a
better quantile regression surface than QR.

\begin{figure*}[b]

\includegraphics{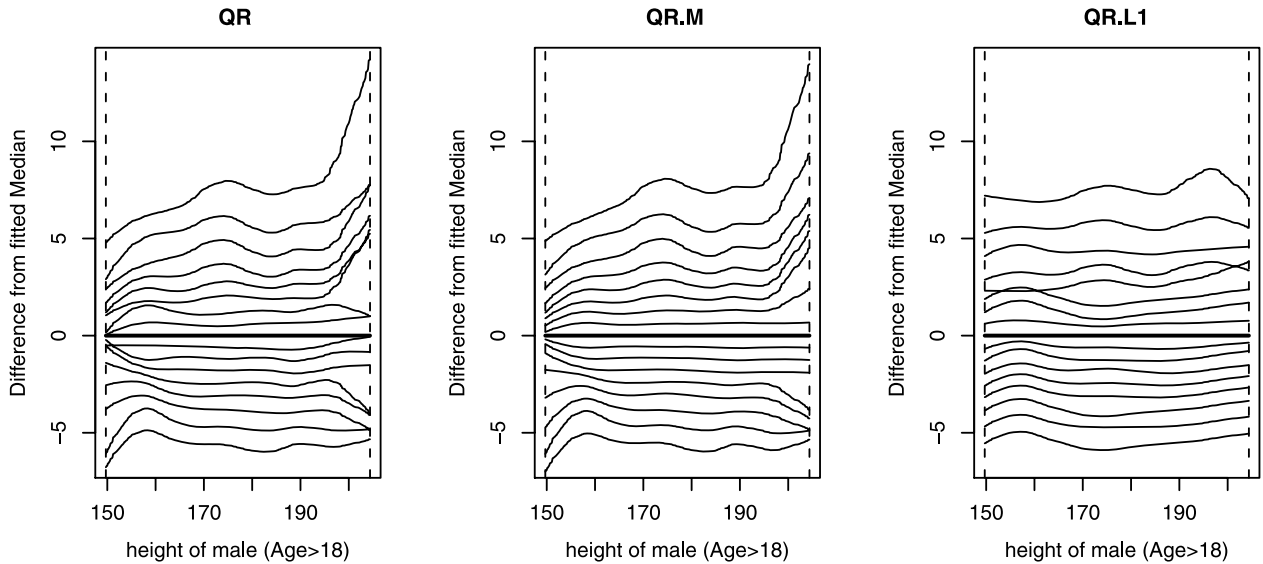}

\caption{Differences between fitted median line and the other fitted
quantiles for standard quantile regression (QR), modified quantile
regression (QR.M), and $\ell_1$ penalized quantile regression (QR.L1)
for the NHANES data.
The dashed lines are the minimum and maximum of the observed heights.}
\label{fig:NHANES_diff}
\end{figure*}

Modified quantile regression has an additional advantage which
is apparent for small and large heights. The standard quantile
regression fits show several crossings of estimated quantiles,
while crossing behavior
is reduced considerably with modified quantile regression.
Crossed quantiles correspond to\break a~claim that a lower quantile
lies above a higher quantile, contradicting the laws of probability.
Figure \ref{fig:NHANES_diff} shows this behavior.
Fixes for this behavior have been proposed (e.g., \citep{He:1997}), but
we consider it desirable to lessen crossing without any explicit
fix. The reduction in crossing holds up across other data sets that we have
examined and with regression models that differ in their details.

%
\begin{table}
\caption{Mean and (1000 times standard deviation) of CV
scores at selected quantiles based on 500 replicates from NHANES data}\label{tab:NHANES}
\begin{tabular*}{\columnwidth}{@{\extracolsep{\fill}}lccc@{}}
\hline
\textbf{Method} & $\bolds{q=0.25}$ & $\bolds{q=0.5}$ & \multicolumn{1}{c@{}}{$\bolds{q=0.9}$} \\
\hline
QR & 1.5040 (0.6105) & 2.0405 (0.7272) & 1.1267 (1.0714)\\
QR.M & 1.5039 (0.5855) & 2.0402 (0.6576) & 1.1263 (1.0030)\\
QR.L1 & 1.5039 (0.8963) & 2.0393 (0.5140) & 1.1289 (0.8569)\\
\hline
\end{tabular*}
\end{table}

In addition, we compare both methods with $\ell_1$ parameter-penalized quantile
regression (QR.L1),\break
where the estimator $\hat{\beta}$ is defined as the minimizer~of
\[
\sum_{i=1}^n\rho_q(y_i-x_i^\top\beta) + \lambda_{\beta}\sum
_{j=1}^p|\beta_j|.
\]
The \texttt{rq.fit.lasso} function in the \texttt{quantreg} R package was
used for QR.L1.\vadjust{\goodbreak}
Keeping the same split of data into 90\% of training and 10\% of
testing for each replicate, we have chosen $\lambda_\beta$
among 100 candidate values by 9-fold cross-validation.
The results are in Table \ref{tab:NHANES}.

The effect of parameter penalization differs from modification of
the loss function. Figure \ref{fig:NHANES_diff} illustrates the difference.
The quantiles estimated under QR.L1 (with $\lambda_\beta$ chosen by
10-fold cross-validation) show less variation across $x$
relative to the fitted median line, due to the shrinkage of each $\beta_j$
toward 0. This effect is more visible for large quantiles.
Such nondifferential penalty can degrade performance, unless
the parameters are of comparable size.
This adverse effect is numerically evidenced in the large $\mathrm{CV}$ score of
QR.L1 for $q=0.9$ in Table \ref{tab:NHANES}.
For $q=0.25$ and $0.5$, QR.L1 yields similar results to the other two
methods in terms of the $\mathrm{CV}$ scores.

\subsection{Analysis of Language Data}
\label{subsec:langdata}

\citet{balota04} conducted an extensive lexical decision experiment
in which subjects were asked to identify whether a string of letters was
an English word or a nonword. The words were monosyllabic, and the
nonwords were constructed to closely resemble words on a number of
linguistic dimensions. Two groups were studied---college students and
older adults. The data consist of response times by word, averaged over
the thirty subjects in each group. For each word, a number of
covariates was recorded. Goals\vadjust{\goodbreak} of the experiment include determining
which features of a word (i.e., covariates) affect response time, and
whether the active features affect response time in the same fashion
for college students and older adults. The authors make a case for the
need to conduct and analyze studies with regression techniques in mind,
rather than simpler ANOVA techniques.

\citet{baayen} conducted an extensive analysis of a slightly modified data
set which is available in his \texttt{languageR} package. In his
analysis, he creates and selects variables to include in a regression
model, addresses issues of nonlinearity, collinearity and interaction,
and removes selected cases as being influential and/or outlying. He
trims a total of $87$ of the $4568$ cases. The resulting model, based
on ``typical'' words, is used to address issues of linguistic importance.
It includes seventeen basic covariates which enter the model as linear
terms, a nonlinear term for the written frequency of a word (fit as a
restricted cubic spline with five knots), and an interaction term between
the age group and the (nonlinear) written frequency of the word.

We consider two sets of potential covariates for the model. The small set
consists of Baayen's $17$ basic covariates and three additional covariates
representing a squared term for written frequency and the interaction
between age group and the linear and squared terms for written frequency.
Age group has been coded as $\pm1$ for the interactions. The large set
augments these covariates with nine additional covariates that were not
included in Baayen's final model. Baayen excluded some of these covariates
for a lack of significance, others because of collinearity.

To investigate the performance of the LASSO and robust LASSO, a simulation
study was conducted on the $4568$ cases in the data set. For a single
replicate in the simulation, the data were partitioned into a training
data set and a test data set. The various methods were fit to the training
data, with evaluation conducted on the test data. The criteria for
evaluation were sum of squared differences between the fitted and observed
responses, either over all cases in the test data or over the test data
with the cases identified by Baayen as outliers removed. We refer to these
criteria as predictive mean squared error ($\mathrm{PMSE}$).

The simulation investigated several factors, including the amount of
training data (10\% of the full data, 20\%, 30\%, etc.), the regularization
parameter $\lambda_{\gamma} = k \hat{\sigma}$, and
the method used to select the model.
Three methods were used to
select the model (i.e., the fraction\vadjust{\goodbreak} of the distance along the solution
path): minimum $C_p$, generalized cross-validation,
and 10-fold cross-validation on the training data.

\begin{figure*}
\centering
\begin{tabular}{@{}cc@{}}

\includegraphics{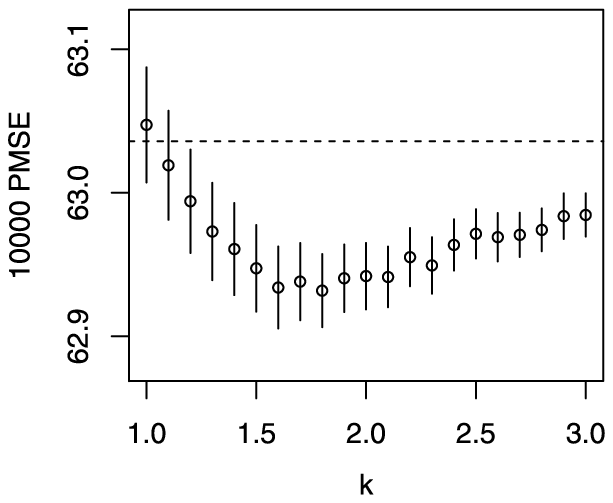}
 & \includegraphics{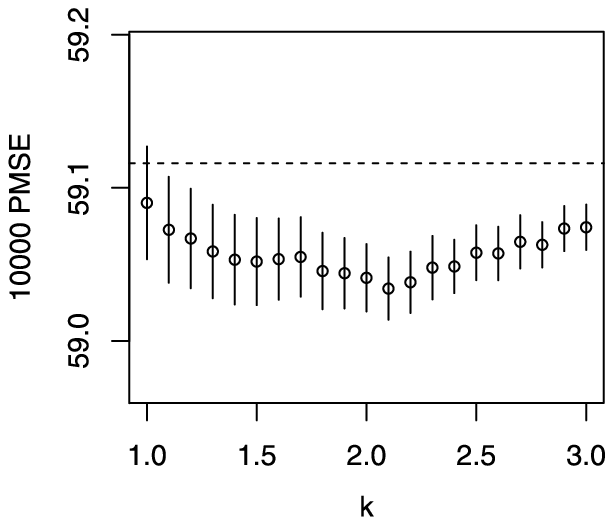}\\
\footnotesize{(a)} & \footnotesize{(b)}
\end{tabular}
\caption{Predictive mean squared error (\textit{PMSE}) for the test data in the
simulation study, after removal of cases identified by Baayen
as outliers.
The horizontal
line is the mean \textit{PMSE} for the LASSO while the points represent
the mean of \textit{PMSE}s for the robust LASSO. The vertical lines have
the width of
approximate 95\% confidence intervals for the difference
in mean \textit{PMSE} under the LASSO and robust LASSO.
Panel \textup{(a)} presents results
for the small set of covariates and panel \textup{(b)} presents results for
the large set of covariates.}
\label{fig:LangSimPMSE}
\end{figure*}

The results of a $300$ replicate simulation show a~convincing benefit to
use of the robust LASSO. The benefit of the robust LASSO is most apparent
when~$k$ is in the ``sweet spot'' ranging from $1.4$ or so to well
above $2.0$. As
expected, for very small $k$ (near~1), the robust LASSO may not perform
as well as the LASSO. The reduction in $\mathrm{PMSE}$ for moderate values of $k$,
both absolute and percent, is slightly larger when the evaluation is
conducted after outliers (as identified by Baayen---not by the fitted model)
have been dropped from the test data set. The benefit is largest for small
training data sets and decreases as the size of the training data set
increases. For large training data sets (e.g., 90\% of the data), little
test data remains for calculation of $\mathrm{PMSE}$ and the evaluation is less
stable. These patterns were apparent over all three methods of model
selection. Figure~\ref{fig:LangSimPMSE}
shows the results for a training sample size of $1827$
cases (40\% of the data), with model selected by cross-validation, for a
variety of values of $k$. The $\mathrm{PMSE}$ for the robust LASSO dips below the
mean $\mathrm{PMSE}$ for the LASSO for a wide range of $k$. The figure also presents
95\% confidence intervals, based on the $300$ replicates in the simulation,
for the difference between mean $\mathrm{PMSE}$ under the robust LASSO and the LASSO.
The intervals are indicated by the vertical lines, and statistical
significance is indicated where the lines do not overlap the mean
$\mathrm{PMSE}$ under the LASSO. The narrowing of the intervals is a consequence
of the greater similarity of LASSO and robust LASSO fits as the bending
constant increases.
The patterns just described hold for both the small set of
covariates and the large set
of covariates.


%

%
\begin{figure*}[b]
\centering
\begin{tabular}{@{}cc@{}}

\includegraphics{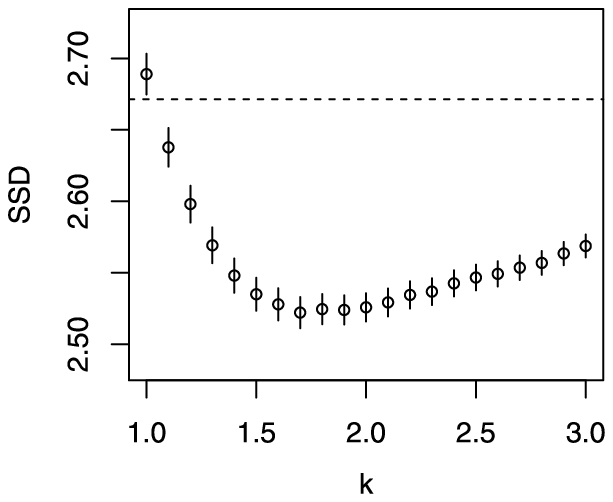}
 & \includegraphics{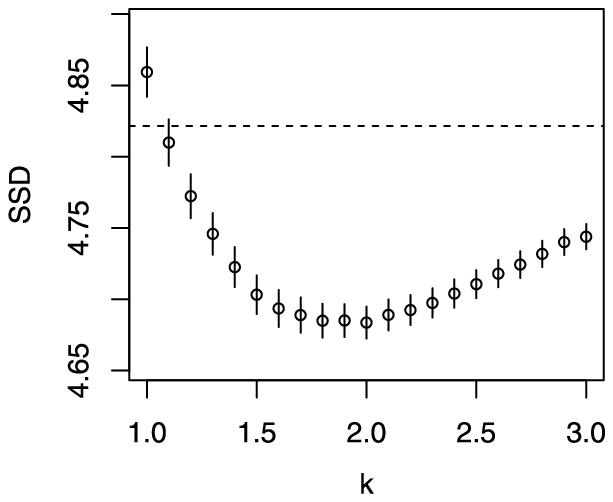}\\
\footnotesize{(a)} & \footnotesize{(b)}
\end{tabular}
\caption{Sum of squared deviations (\textit{SSD}) from Baayen's fits in the simulation
study. The horizontal
line is the mean \textit{SSD} for the LASSO while the points represent
the mean of SSDs for the robust LASSO. The vertical lines have
the width of
approximate 95\% confidence intervals for the difference in mean \textit{SSD}
under the LASSO and robust LASSO.
Panel \textup{(a)} presents results
for the small set of covariates and panel \textup{(b)} presents results for
the large set of covariates.}
\label{fig:LangSimSSD}
\end{figure*}

In addition to using the test data as a target, we studied how well the
two methods could reproduce Baayen's expert fit. This makes a good target
for inference, as there is evidence that humans can produce a better fit
than automated methods (\citep{Yu:AOAS2011}). Taking a fitted surface as a
target allows us to remove the noise inherent in data-based out-of-sample
evaluations.\break The results from a $5000$ replicate simulation study with a
training sample size of $400$ appear in Figure~\ref{fig:LangSimSSD}.
The criterion is sum
of squared deviations ($\mathrm{SSD}$) between the (robust) LASSO fit and Baayen's
fit, with the sum taken over only those covariate values contained in the
test data set. The results presented here are for models
selected with the minimum $C_p$ criterion.
The robust LASSO outperforms the LASSO over a wide range of values for $k$
for both the small and large sets of covariates.

Figures \ref{fig:LangSimPMSE} and \ref{fig:LangSimSSD} reveal an
interesting difference across targets in the behavior of the
small and large sets of covariates. When the target is an
expert fit, as in the second study, adding covariates not present
in the expert's model to the pool of potential covariates
allows the LASSO and robust LASSO to produce a near-equivalent fit
to the data, but with different coefficients for the regressors.
An examination of the variables present in the fitted models and
their coefficients
uncovers patterns. As an example, the two covariates
``WrittenFrequency'' and ``Familiarity'' appear in nearly all
of the models for both the LASSO and the robust LASSO, while
Baayen includes only ``WrittenFrequency'' in his model, and these
covariate(s) have negative coefficients.
Subjects are able to decide that a familiar
word is a~word more quickly (and more accurately) than an
unfamiliar word. Although there seems to be no debate on
whether this conceptual effect of similarity exists, there
are a variety of viewpoints on how to best capture the effect.
Regularization methods allow one to include a suite of
covariates to address a single conceptual effect, and this
produces a difference between the LASSO and robust LASSO fits
on one hand and a least-squares, variable-selection style fit
on the other hand. The end result is that the regularized
fits with the large set
of covariates show greater departures from Baayen's fit than
do regularized fits with the small set of covariates.
In contrast, under the data-based target
of the first study, the large set of covariates results in a
smaller $\mathrm{PMSE}$.

\section{Discussion}
\label{sec:disc}

In the preceding sections, we have laid out an approach to
modifying modeling procedures. The approach is based
on the creation of case-specific covariates which are then
regularized. With appropriate choices of
penalty terms, the addition of these covariates allows us to
robustify those procedures which lack robustness and also allows
us to improve the efficiency of procedures which are very robust,
but not particularly efficient. The method is fully compatible
with regularized estimation methods. In this case, the case-specific
covariates are merely included as part of the regularization.
The techniques are easy to
implement, as they often require little modification of existing
software. In some cases, there is no need for modification of
software, as one merely feeds a modified data set into
existing routines.

The motivation behind this work is a desire to move relatively
automated modeling procedures
in the direction of traditional data analysis
(e.g., Weisberg, \citeyear{weisberg:lars}).
An important component of this type of analysis
is the ability to take different looks at a~data set. These
different looks may suggest creation of new variates
and differential handling of individual cases or groups of cases.
Robust methods allow us to take such a look, even when data sets
are large. Coupling robust regression techniques with the ability
to examine an entire solution path provides a~sharper view of the
impact of unusual cases on the analysis.
A second motivation for the work is the desire to improve robust,
relatively nonparametric methods. This is accomplished by introducing
case-specific parameters in a controlled fashion whereby the finite sample
performance of estimators is improved.

The perspective provided by this work suggests several directions for
future research. Adaptive penalties, whether used for robustness or
efficiency, can be designed to satisfy specified invariances.
The asymmetric $\ell_2$ penalty for modified quantile regression was
designed to satisfy a specified invariance. For a locally constant
residual density, it keeps the $0$ of the $\psi_q^\gamma$ function
invariant as the width of the interval of adjustment varies.
Specific, alternative forms of invariance for quantile regression are
suggested by\vadjust{\goodbreak} consideration of parametric driving forms for the residual
distribution. A motivating parametric model, coupled with invariance
of the $0$ of the $\psi_q^\gamma$ function to the size of the penalty
term $\lambda_\gamma$, yields a~path of penalties. Increasing the size
of the covari\-ate-specific penalty at an appropriate rate leads to
asymptotic equivalence with the quantile regression estimator.
This allows one to fit the model nonparametrically
while tapping into an approximate parametric form
to enhance finite sample performance. Similarly, when case-specific
penalties are applied to a model such as the generalized linear model,
the asymmetry of the likelihood, coupled with invariance, suggests an
asymmetric form for the $\ell_1$ penalty used to enhance robustness
of inference.

Following development of the technique for quantile regression, one can
apply the adaptive loss para\-digm for model assessment and selection.
For example, in cross-validation, a summary of a model's fit is computed
as an out-of-sample estimate of empirical risk and the evaluation is
used for choosing the model (parameter value) with the smallest estimated
risk. For model averaging, estimated risks are converted into weights which
are then attached to model-specific predictions that are then combined
to yield an overall prediction. The use of modified loss functions for
estimation of risks is expected to improve stability and efficiency in
model evaluation and selection.\vspace*{-1pt}

\begin{appendix}
\section*{Appendix}\label{app}
Proof of Theorem \ref{thm:reg_qtl}.
Let $u_i:=y_i-x_i^\top\beta(q)$ and consider the objective function
%
\renewcommand{\theequation}{\arabic{equation}}
\setcounter{equation}{12}
\begin{equation}
\quad Z_n^\gamma(\delta):=
\sum^{n}_{i=1}\bigl\{\rho_q^\gamma\bigl(u_i-x_i^\top\delta/\sqrt{n}\bigr)-\rho_q(u_i)\bigr\}.
\end{equation}
Note that $Z_n^\gamma(\delta)$ is minimized at $\hat\delta_n:=
\sqrt{n}(\hat{\beta}^\gamma_{n}-\beta(q))$, and
the limiting distribution of $\hat\delta_n$ is determined by
the limiting behavior of $Z_n^\gamma(\delta)$. To study the limit
of $Z_n^\gamma(\delta)$, decompose $Z_n^\gamma(\delta)$ as
\begin{eqnarray*}
Z_n^\gamma(\delta)& = &
\sum^{n}_{i=1}\bigl\{\rho_q^\gamma\bigl(u_i-x_i^\top\delta/\sqrt{n}\bigr)-
\rho_q\bigl(u_i-x_i^\top\delta/\sqrt{n}\bigr)\bigr\}
\\
&&{}+\sum^{n}_{i=1}\bigl\{\rho_q\bigl(u_i-x_i^\top\delta/\sqrt{n}\bigr)-\rho_q(u_i)\bigr\}\\
&=&\sum^{n}_{i=1}\bigl\{\rho_q^\gamma\bigl(u_i-x_i^\top\delta/\sqrt{n}\bigr)-
\rho_q\bigl(u_i-x_i^\top\delta/\sqrt{n}\bigr)\bigr\}\\
&&{}+Z_n(\delta),
\end{eqnarray*}
where $Z_n(\delta)
:=\sum^{n}_{i=1}\{\rho_q(u_i-x_i^\top\delta/\sqrt{n})-\rho_q(u_i)\}$.
By showing that the first sum converges to zero in\vadjust{\goodbreak} probability
up to a sequence of constants that do not depend on $\delta$,
we will establish the asymptotic equivalence
of $\hat{\beta}^\gamma_{n}$ to $\hat{\beta}_{n}$.

Given $\lambda_\gamma=cn^\alpha$, first observe that
\begin{eqnarray*}
&&\hspace*{-5pt}E \bigl\{ \rho_q^\gamma\bigl(u_i-x_i^\top\delta/\sqrt{n}\bigr)-
\rho_q\bigl(u_i-x_i^\top\delta/\sqrt{n}\bigr)\bigr\}\\
&&\hspace*{-5pt}\qquad{}+ q(1-q)/2\lambda_\gamma\\
&&\hspace*{-5pt}\quad=\int^{{(1-q)}/{\lambda_\gamma}+x_i^\top\delta/\sqrt{n}}_{x_i^\top
\delta/\sqrt{n}}
\biggl(\frac{\lambda_\gamma}{2} \frac{q}{1-q}
\biggl(u-\frac{x_i^\top\delta}{\sqrt{n}}\biggr)^2
\\
&&\hspace*{82pt}\qquad{}-q\biggl(u-\frac{x_i^\top\delta}{\sqrt{n}}\biggr)
+ \frac{q(1-q)}{2\lambda_\gamma}\biggr)\\
&&\hspace*{101pt}\cdot f_i(\xi_i+u)\,du\\
&&\hspace*{-5pt}\qquad{}+\int^{x_i^\top\delta/\sqrt{n}}_{-q/{\lambda_\gamma}+x_i^\top\delta
/\sqrt{n}}
\biggl(\frac{\lambda_\gamma}{2} \frac{1-q}{q}
\biggl(u-\frac{x_i^\top\delta}{\sqrt{n}}\biggr)^2
\\
&&\hspace*{79pt}\qquad{}-(q-1)\biggl(u-\frac{x_i^\top\delta}{\sqrt{n}}\biggr)
\\
&&\hspace*{128pt}\qquad{}
+ \frac{q(1-q)}{2\lambda_\gamma}\biggr)\\
&&\hspace*{96pt}{}\cdot f_i(\xi_i+u)\,du\\
&&\quad=\int^{{(1-q)}/{\lambda_\gamma}+x_i^\top\delta/\sqrt{n}}_{x_i^\top
\delta/\sqrt{n}}
\frac{\lambda_\gamma}{2} \frac{q}{1-q}\\
&&\hspace*{79pt}\qquad{}\cdot\biggl(u-\frac{x_i^\top\delta}{\sqrt{n}}-\frac{1-q}{\lambda_\gamma}\biggr)^2\\
&&\hspace*{79pt}\qquad{}\cdot f_i(\xi_i+u)\,du\\
&&\qquad{}+\int^{x_i^\top\delta/\sqrt{n}}_{-q/{\lambda_\gamma}+x_i^\top\delta
/\sqrt{n}}
\frac{\lambda_\gamma}{2} \frac{1-q}{q}
\biggl(u-\frac{x_i^\top\delta}{\sqrt{n}}+\frac{q}{\lambda_\gamma}\biggr)^2
\\
&&\hspace*{79pt}\qquad{}\cdot f_i(\xi_i+u)\,du.
\end{eqnarray*}
Using a first-order Taylor expansion of $f_i$ at $\xi_i$ from the
condition (C-2) and the expression above, we have
\begin{eqnarray*}
&&E \sum^{n}_{i=1}\bigl\{\rho_q^\gamma\bigl(u_i-x_i^\top\delta/\sqrt{n}\bigr)-
\rho_q\bigl(u_i-x_i^\top\delta/\sqrt{n}\bigr)\bigr\} \\
&&\qquad{}+ nq(1-q)/2\lambda_\gamma\\
&&\quad= \frac{q(1-q)}{6c^2n^{2\alpha}}\sum^{n}_{i=1}f_i(\xi_i) +
\frac{q(1-q)}{6c^2n^{2\alpha}}
\sum^{n}_{i=1}\frac{f'_i(\xi_i)x_i^\top\delta}{\sqrt{n}}
\\
&&\qquad{}+o(n^{-2\alpha+1/2}).
\end{eqnarray*}
Note that $\sum^{n}_{i=1}f'_i(\xi_i)x_i^\top\delta/\sqrt{n}=O(\sqrt
{n})$ as
$f'_i(\xi_i)$, $i=1,\ldots,n$, are uniformly bounded from the
condition (C-2), and
$|x_i^\top\delta| \le\|x_i\|_2\|\delta\|_2\le
(\|x_i\|_2^2+\|\delta\|_2^2)/2$ while
$\sum^{n}_{i=1} \|x_i\|_2^2=O(n)$ from\vadjust{\goodbreak} the condition \mbox{(C-3)}.
Taking
$C_n:= -{q(1-q)}/ {(2cn^{\alpha-1})}+
{q(1-q)}/\break {(6c^2n^{2\alpha})}\sum^{n}_{i=1}f_i(\xi_i)$,
we have that
\begin{eqnarray*}
&&E \sum^{n}_{i=1}\bigl\{\rho_q^\gamma\bigl(u_i-x_i^\top\delta/\sqrt{n}\bigr)-
\rho_q\bigl(u_i-x_i^\top\delta/\sqrt{n}\bigr)\bigr\}-C_n \\
&&\quad{}\rightarrow0  \quad\mbox{if }
\alpha> 1/4.
\end{eqnarray*}
Similarly, it can be shown that
\begin{eqnarray*}
&&\operatorname{Var} \sum^{n}_{i=1}\bigl\{\rho_q^\gamma\bigl(u_i-x_i^\top\delta/\sqrt{n}\bigr)-
\rho_q\bigl(u_i-x_i^\top\delta/\sqrt{n}\bigr)\bigr\}\\
&&\quad= \sum^{n}_{i=1}\frac{q^2(1-q)^2f_i(\xi_i)}{20c^3n^{3\alpha}}
+o(n^{-3\alpha+1})\\
&&\qquad\rightarrow0  \quad\mbox{if } \alpha> 1/3.
\end{eqnarray*}
Thus, if $\alpha$ $>1/3$,
\begin{eqnarray*}
&&\sum^{n}_{i=1}\rho_q^\gamma\bigl(u_i-x_i^\top\delta/\sqrt{n}\bigr)-
\rho_q\bigl(u_i-x_i^\top\delta/\sqrt{n}\bigr) - C_n \\
&&\quad\rightarrow0
\quad\mbox{in probability}.
\end{eqnarray*}
This implies that the limiting behavior of $Z_n^\gamma(\delta)$ is
the same as
that of $Z_n(\delta)$. From the proof of Theorem 4.1 in \citet{Koenker:2005},
$Z_n(\delta) \stackrel d \rightarrow-\delta^\top W
+\frac{1}{2}\delta^\top D_1\delta$,
where $W$ $\sim N(0,q(1-q)D_0)$. By the convexity argument in \citet
{Koenker:2005}
(see also \citep{pollard:1991}; \citep{hjort:pollard}; \citep{Knight:1998}),
$\hat\delta_n$,
the minimizer of $Z_n^\gamma(\delta)$,
converges to $\hat\delta_0:=D_1^{-1}W$, the unique minimizer of
$-\delta^\top W+\frac{1}{2}\delta^\top D_1\delta$ in distribution.
This completes the proof.
\end{appendix}

\section*{Acknowledgments}
We thank the Editor, Associate Editor and referees for their
thoughtful comments which helped us improve the presentation of this paper.
This research was supported in part by NSA Grant H98230-10-1-0202.

%

\end{document}